\documentclass[structabstract]{aa}  
\usepackage{graphicx}
\usepackage{txfonts}
\usepackage{natbib}
\usepackage{color}
\newcommand{\comment}[1]{}
\newcommand{\vel}{\upsilon}

\begin{document}
   \title{Three-dimensional simulations of near-surface convection in main-sequence stars}
   \subtitle{IV. Effect of small-scale magnetic flux concentrations on centre-to-limb variation and spectral lines}

   \author{B. Beeck\inst{1}
     	  \and
	  M. Sch\"ussler\inst{1}
          \and
          R.~H. Cameron\inst{1}
	  \and
	  A. Reiners\inst{2}
          }

   \institute{Max-Planck-Institut f\"ur Sonnensystemforschung,
              Justus-von-Liebig-Weg 3, 37077 G\"ottingen, Germany\\
         \and
              Institut f\"ur Astrophysik, 
	      Universit\"at G\"ottingen, 
	      Friedrich-Hund-Platz 1,
	      37077 G\"ottingen, Germany\\
             }

   \date{Received 12 February 2015 / Accepted 20 April 2015}
   \titlerunning{3D simulations of near-surface convection in main-sequence
     stars. IV.}

\abstract
{Magnetic fields affect the local structure of the
  photosphere of stars. They can considerably influence the radiative
  properties near the optical surface, flow velocities, and
  the temperature and pressure profiles. This has an impact on observables
  such as limb darkening and spectral line profiles.}
{We aim at understanding qualitatively the influence of small
  magnetic flux concentrations in unipolar plage regions on the centre-to-limb
  variation of the intensity and its contrast and on the shape of
  spectral line profiles in cool main-sequence stars.}
{We analyse the bolometric and continuum intensity and its angular dependence
  of 24 radiative magnetohydrodynamic simulations of the
  near-surface layers of main-sequence stars with six different sets of
  stellar parameters (spectral types F to early M) and four different average
  magnetic field strengths (including the non-magnetic case). We also
  calculated disc-integrated profiles of three spectral lines.}
{The small magnetic flux concentrations formed in the magnetic runs of
  simulations have a considerable impact on the intensity and its
  centre-to-limb variation. In some cases, the difference in limb darkening
  between magnetic and non-magnetic runs is larger than the difference
  between the spectral types. Spectral lines are not only broadened owing
  to the Zeeman effect, but are also strongly affected by the modified
  thermodynamical structure and flow patterns. This indirect
  magnetic impact on the line profiles is often bigger than that of the Zeeman effect.}
{The effects of the magnetic field on the radiation leaving the star can be
  considerable and is not restricted to spectral line broadening and
  polarisation by the Zeeman effect. The inhomogeneous structure of the
  magnetic field on small length scales and its impact on (and spatial
  correlation with) the local thermodynamical structure and the flow field
  near the surface influence the measurement of the global field
  properties and stellar parameters. These effects need to be taken into
  account in the interpretation of observations.}


   \maketitle
%
\section{Introduction}
%
%
Many cool main-sequence stars have substantial surface magnetic fields
\citep[see][for a review]{Ansgar12}. These fields are believed to be generated
by a hydrodynamic dynamo in the deeper parts of the convection zone and are
transported to the surface layers by buoyancy and convective flows
\citep[see][for a review]{Charbonneau10}. Knowledge of the global geometry and
strength of stellar surface magnetic fields and their dependence on stellar
parameters is important in order to understand the underlying
nature of the dynamo processes in stars of different internal structure
\citep[e.\,g.][]{Donati11}. Moreover, the magnetic field is an important noise
source for many astronomical observations \citep[such as exo-planet detection, see e.\,g.][]{Jeffersetal2013}, which makes a deeper understanding
of its properties and the impact on observables essential.\par
%
%
The presence of stellar magnetic fields is often indirectly detected by means
of activity indicators (such as chromospheric S-index, X-ray flux) or
photometric variability, which is mainly attributed to the rotational
variation of the projected starspot area. Fast rotators in general tend to be
more magnetically active \citep[see][and references
  therein]{Ansgar2014}. Moreover, at least some stars were found to
exhibit activity cycles analogous to the 11-year solar activity cycle
\citep{MtWilson}.\par
%
%
In some cases, it is possible to measure the surface magnetic field by
spectroscopic or spectropolarimetric means, which exploit the action of the
Zeeman effect on spectral line profiles in the integrated (Stokes-$I$) and
circularly polarised (Stokes-$V$) light. While line broadening in Stokes-$I$
gives an estimate of the unsigned average magnetic field of a star, Stokes-$V$
measurements are sensitive to the polarity of the field so that they pick up
only the uncancelled net flux (signed average field). For main-sequence stars
of spectral types F, G, and K, there exist only a few (Stokes-$I$) magnetic field
measurements, as the magnetic line broadening owing to the Zeeman effect is
usually much smaller than the rotational broadening in these stars
\citep[see][and references therein]{Ansgar12}. For M dwarfs, such measurements
are somewhat more feasible as the rotational velocities required for
the generation of measurable magnetic fields are smaller. Some M dwarfs are
found to be completely covered with a magnetic field in the kG regime, with
local values of up to $\sim 7\,\mathrm{kG}$ \citep{Denis14}.\par
For rapidly rotating stars, it is possible to infer the large-scale geometry
of the magnetic field with spectropolarimetry: Zeeman-Doppler imaging (ZDI)
exploits the latitude-dependent time dependence of the Zeeman-effect signature
in polarised light caused by the Doppler effect due to the stellar
rotation. ZDI reveals large-scale magnetic geometries which are often quite
unlike the solar one \citep{Morin10}.\par
%
%
Owing to their interaction with the convective flows, stellar surface magnetic
fields are expected to be highly structured and inhomogeneous on spatial
scales as small as the scale of the convection, as it is known for the Sun:
the magnetic flux at the solar surface is mainly concentrated in sunspots,
pores, and smaller magnetic structures in the intergranular lanes. In these
magnetic flux concentrations, the magnetic forces cause local modifications of
the thermodynamical structure and the convective flows in the photospheric
layers, which result, e.\,g. in a modulation of the local intensity emerging
from magnetic flux concentrations. These modifications should be taken into
account in spectroscopic and spectropolarimetric measurements of stellar
magnetic fields as they will have a considerable influence on the shape of the
local spectral line profiles. \citet{Basri1990} and \citet{SaSo92} analysed
the significant impact of multicomponent atmospheres and magnetic field
gradients on spectral lines and the consequences for stellar magnetic field
measurements. More recently, \citet{Rosenetal12} showed that Zeeman-Doppler
imaging of the magnetic field for a star with cool (dark) magnetic spots can
fail if one does not simultaneously reconstruct the
temperature. Three-dimensional MHD simulations of the near-surface convection
as presented in this paper series provide a realistic and self-consistent
structure of the magnetic field and stellar atmosphere and consequently can
help to considerably improve stellar magnetic field measurements. \par
%
%
In \citet[][hereafter Paper~I]{paper1}, we presented non-magnetic
3D simulations of six main-sequence stars (spectral types F3V
to M2V). In \citet[][hereafter Paper~II]{paper2}, we analysed the impact of
the convective flows and the surface structure caused by them on the limb
darkening and on the shapes of the profiles of three spectral lines. In
\citet[hereafter Paper~III]{paper3} we investigated a grid of MHD simulations
with the same stellar parameters as the non-magnetic simulations analysed in
Paper~I and Paper~II. The magnetic simulation runs include a unipolar magnetic
field with an average field strength of up to 500\,G. In this paper, we
analyse the impact of the magnetic field on the intensity field and spectral
line profiles in the grid of simulations introduced in Paper~I and Paper~III.

\section{Methods}
\begin{table*}
\centering
\caption{Disc-centre ($\mu=1$) intensities.}\label{tab:I}
\begin{tabular}{lrr@{\,$\pm$\,}lr@{.}lr@{.}lr@{.}lr@{.}l}\hline\hline
Run (SpT - $B_0$)\,$^{\mathrm{a}}$ & $\log g$\,[cgs] & \multicolumn{2}{c}{$T_{\mathrm{eff}}$\,[K]} &\multicolumn{2}{c}{$\langle I_{\mathrm{bol}}/I_{\mathrm{bol},\odot, \mathrm{hydro.}}\rangle$} & \multicolumn{2}{c}{$\sqrt{\langle I_{\mathrm{bol}}^2\rangle}/\langle I_{\mathrm{bol}}\rangle$} & \multicolumn{2}{c}{$\min(I_{\mathrm{bol}})/\langle I_{\mathrm{bol}} \rangle\,^{\mathrm{b}}$} & \multicolumn{2}{c}{$\max(I_{\mathrm{bol}})/\langle I_{\mathrm{bol}} \rangle\,^{\mathrm{b}}$}\\\hline
F3V - hydro.\,$^{\mathrm{c}}$ & 4.301 & 6893 & 6 & \quad 1&992  & \qquad 0&206  & \qquad 0&279 & \qquad 2&031 \\  
F3V - 20\,G  & 4.301 & 6885 & 6 & \quad 1&989  & \qquad 0&203  & \qquad 0&356 & \qquad 2&238 \\ 
F3V - 100\,G & 4.301 & 6911 & 8 & \quad 1&994  & \qquad 0&201  & \qquad 0&392 & \qquad 2&682 \\
F3V - 500\,G & 4.301 & 7003 & 5 & \quad 2&017  & \qquad 0&196  & \qquad 0&345 & \qquad 2&429 \\\hline
G2V - hydro.\,$^{\mathrm{c}}$ & 4.438 & 5764 & 7 & \quad 1&000  & \qquad 0&154  & \qquad 0&533 & \qquad 1&788 \\
G2V - 20\,G  & 4.438 & 5779 & 9 & \quad 1&011  & \qquad 0&152  & \qquad 0&555 & \qquad 2&368 \\
G2V - 100\,G & 4.438 & 5802 & 8 & \quad 1&015  & \qquad 0&153  & \qquad 0&555 & \qquad 2&471 \\
G2V - 500\,G & 4.438 & 5864 & 9 & \quad 1&009  & \qquad 0&150  & \qquad 0&425 & \qquad 2&269 \\\hline
K0V - hydro.\,$^{\mathrm{c}}$ & 4.609 & 4856 & 6 & \quad 0&5055 & \qquad 0&0796 & \qquad 0&728 & \qquad 1&840 \\
K0V - 20\,G  & 4.609 & 4858 & 2 & \quad 0&5060 & \qquad 0&0853 & \qquad 0&739 & \qquad 2&708 \\
K0V - 100\,G & 4.609 & 4878 & 4 & \quad 0&5125 & \qquad 0&0945 & \qquad 0&696 & \qquad 2&632 \\
K0V - 500\,G & 4.609 & 4901 & 2 & \quad 0&5099 & \qquad 0&116  & \qquad 0&505 & \qquad 2&180 \\\hline
K5V - hydro.\,$^{\mathrm{c}}$ & 4.699 & 4368 & 2 & \quad 0&3235 & \qquad 0&0705 & \qquad 0&776 & \qquad 1&687 \\
K5V - 20\,G  & 4.699 & 4376 & 2 & \quad 0&3257 & \qquad 0&0710 & \qquad 0&802 & \qquad 2&141 \\
K5V - 100\,G & 4.699 & 4383 & 3 & \quad 0&3270 & \qquad 0&0734 & \qquad 0&759 & \qquad 2&128 \\
K5V - 500\,G & 4.699 & 4402 & 2 & \quad 0&3269 & \qquad 0&0855 & \qquad 0&624 & \qquad 2&383 \\\hline
M0V - hydro.\,$^{\mathrm{c}}$ & 4.826 & 3905 & 1 & \quad 0&1949 & \qquad 0&0359 & \qquad 0&861 & \qquad 1&203 \\
M0V - 20\,G  & 4.826 & 3907 & 1 & \quad 0&1953 & \qquad 0&0349 & \qquad 0&865 & \qquad 1&592 \\
M0V - 100\,G & 4.826 & 3909 & 1 & \quad 0&1955 & \qquad 0&0350 & \qquad 0&843 & \qquad 1&580 \\
M0V - 500\,G & 4.826 & 3906 & 1 & \quad 0&1929 & \qquad 0&0538 & \qquad 0&733 & \qquad 1&433 \\\hline
M2V - hydro.\,$^{\mathrm{c}}$ & 4.826 & 3689 & 1 & \quad 0&1525 & \qquad 0&0219 & \qquad 0&904 & \qquad 1&055 \\
M2V - 20\,G  & 4.826 & 3691 & 1 & \quad 0&1526 & \qquad 0&0218 & \qquad 0&902 & \qquad 1&283 \\
M2V - 100\,G & 4.826 & 3692 & 1 & \quad 0&1528 & \qquad 0&0232 & \qquad 0&865 & \qquad 1&262 \\
M2V - 500\,G & 4.826 & 3679 & 1 & \quad 0&1498 & \qquad 0&0519 & \qquad 0&787 & \qquad 1&189 \\\hline
\end{tabular}
\begin{list}{}{}
\item[$^{\mathrm{a}}$] SpT: spectral type; $B_0$: average vertical magnetic field strength
\item[$^{\mathrm{b}}$] global extrema occurring during a time-resolved sample of intensity snapshots spanning several minutes.
\item[$^{\mathrm{c}}$] non-magnetic (i.\,e. {\it hydrodynamic}) simulation.
\end{list}
\end{table*}
The grid of six non-magnetic and 18 magnetic simulations was run with the 3D
radiative RMHD code \texttt{MURaM} \citep{MURaM1, MURaM2}. It comprises six
sets of stellar parameters ($T_{\mathrm{eff}}$, $\log g$), which roughly
follow the (zero-age) main-sequence and correspond to stars of spectral types
F3V, G2V, K0V, K5V, M0V, and M2V (see Paper~I for more details). The
metallicity was assumed to be solar \citep[abundances by][]{AnGr89} in all
cases. The code uses the OPAL equation of state \citep{opal1,opal2} and
\texttt{ATLAS9} opacities \citep{atlas9}, which were binned into four opacity
bins ($\tau$-sorting applying an averaged 3D atmosphere of the hydrodynamical
run of each simulation as reference; for more details see Paper~I). For each
set of stellar parameters, there is one non-magnetic simulation and three
magnetic simulations with average vertical field strengths of 20, 100, and
500\,G (for more details, see Paper~III). Table~\ref{tab:I} lists the gravitational acceleration and the effective temperature of all 24 simulations considered.\par 
The spectral line synthesis was done with the forward module of the code
\texttt{Spinor} \citep{spinor,spinor2}. Whenever averaged spectral line
profiles are considered, the average is a spatial average over the
$512\times512$ horizontal pixels (of the simulation boxes) as well as a temporal
average over six snapshots which are taken several minutes apart from each
other. These averaged line profiles were calculated for 10 different values of
$\mu$ (cosine of the angle $\vartheta$ between line of sight and stellar
surface). For the disc-integrated line profiles considered, we used a
numerical integration which takes into account differential rotation
according to
\begin{equation}\label{eqn:difrot}
\Omega(\theta)=\Omega_{\mathrm{eq}}\left(1-\alpha\sin^2\theta\right)
\end{equation}
where $\Omega$ is the local angular velocity of the star,
$\Omega_{\mathrm{eq}}$ its value at the stellar equator, $\theta$ the latitude
and $\alpha$ a parameter describing the differential rotation.\par 
In all cases considered here, the star was assumed to be homogeneous on large
scales as we use (six snapshots of) a local-box simulation for the line
synthesis of the disc-integrated line profile. For more details, see Paper~II
and references therein. The simulation snapshots were taken with a cadence of
a few minutes from the relaxed phase of the run (see Paper~III).
\begin{figure*}
\centering
\includegraphics[width=7.1cm]{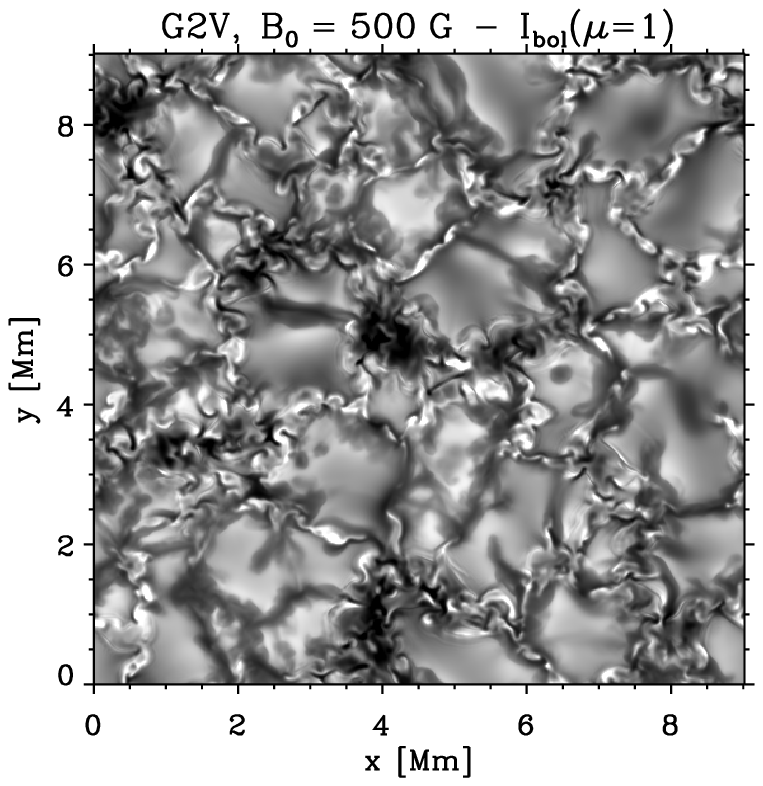}~~\includegraphics[width=7.1cm]{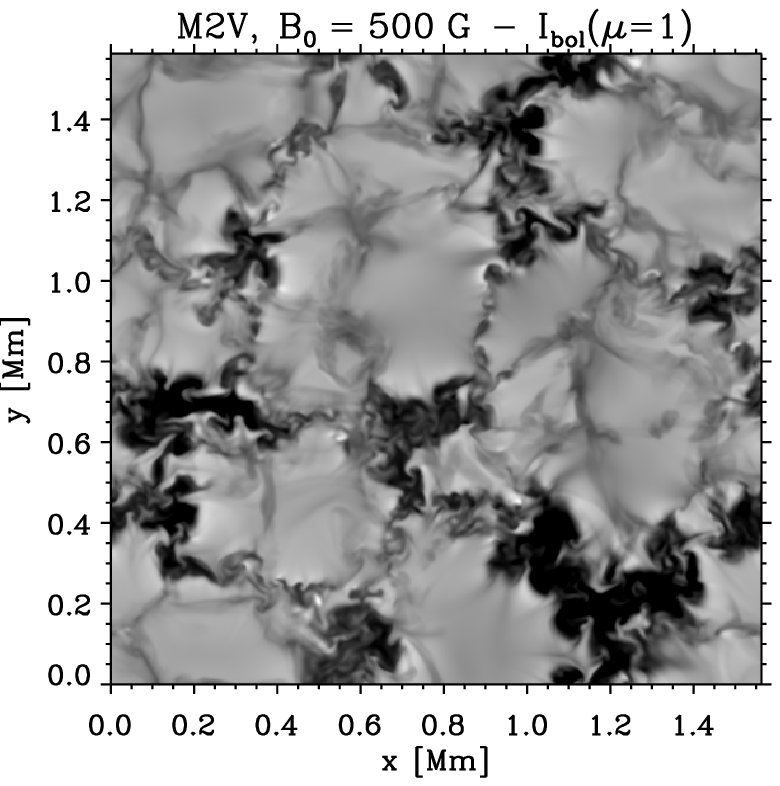}
\caption{Vertically emerging bolometric intensity of a snapshot of the 500\,G runs of the G2V ({\it left}) and M2V ({\it right}) simulations. The grey scale saturates at $\pm 2.5$ standard deviations from the mean intensity. Intensity maps of further simulations are given in Paper I (Fig.~2) and Paper~III (Fig.~1).}\label{fig:I_map}
\end{figure*}

\section{Intensity field}
\subsection{Disc-centre intensity}\label{sec:vert_int}
Figure~\ref{fig:I_map} shows maps of the vertically emerging bolometric
intensity from snapshots of the 500\,G runs of the solar (G2V) and the M2V-star
simulation. As shown in Paper~III, the dark and bright structures in the
intergranular lanes coincide with locations of strong concentrations of
magnetic flux, while the granular upflows are only weakly magnetised. The high
magnetic field strength in the flux concentrations causes a local depression
of the optical surface, which either results in bright structures if the depth
of the depression is comparable or larger than its horizontal extent
(efficient side-wall heating) or dark structures if its depth is small
compared to the horizontal size. The smaller superadiabaticity in M
dwarfs compared to hotter main-sequence stars is responsible for smaller
typical depth-to-radius ratio of these local depressions. Consequently in F-,
G-, and K-type dwarfs, there are only a few dark magnetic structures in the
500\,G runs and none at all in the 100\,G runs whereas the M star photospheres
are dominated by dark magnetic structures at 500\,G and show many of them
already at 100\,G (for more details see Paper~III).\par 

Table~\ref{tab:I} lists the mean bolometric intensities at the disc centre
(i.\,e. $\mu=\cos\vartheta=1$) along with its relative rms contrast and the
extrema of a time-resolved sequence of intensity snapshots spanning several
minutes for each of the 24 simulations considered. The disc-centre intensity
is almost unaffected by the magnetic field in the F-, G-, and K-star
simulations although the radiative flux ($\propto T_{\mathrm{eff}}^4$)
increases significantly owing to magnetic limb brightening (see
Sect.~\ref{sec:clv}). In the M-star simulations, the disc-centre intensity is
reduced in the 500\,G runs with respect to the non-magnetic runs, whereas the
radiative flux is unaffected (M0V) or only marginally reduced (M2V).\par
\begin{figure*}
\centering
\includegraphics[width=7.1cm]{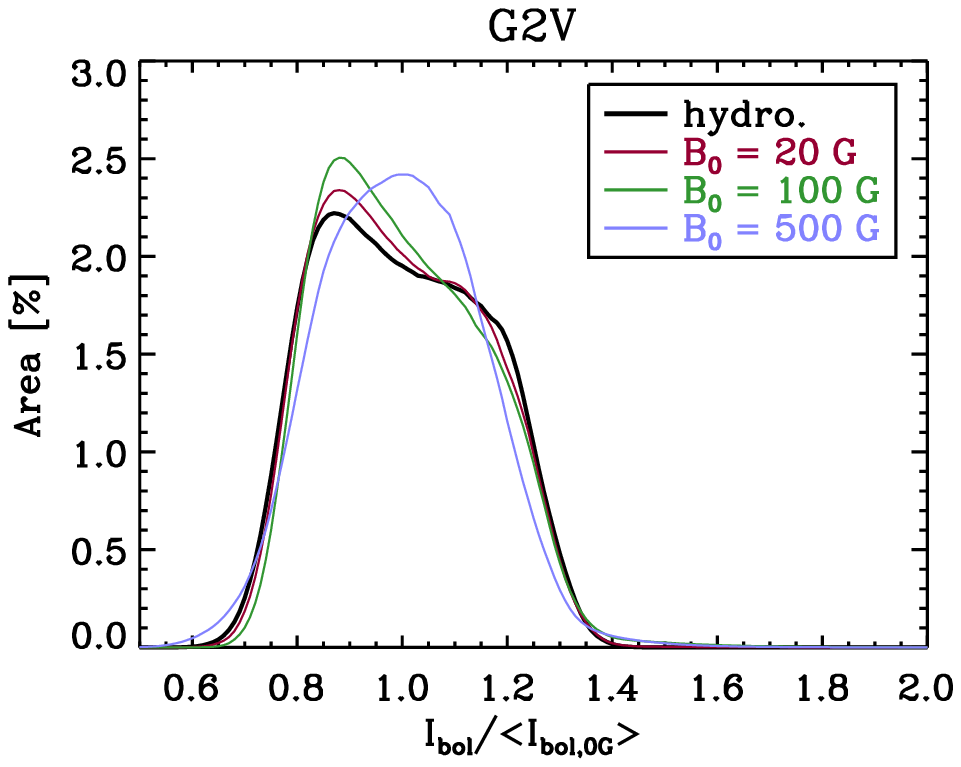}~\includegraphics[width=7.1cm]{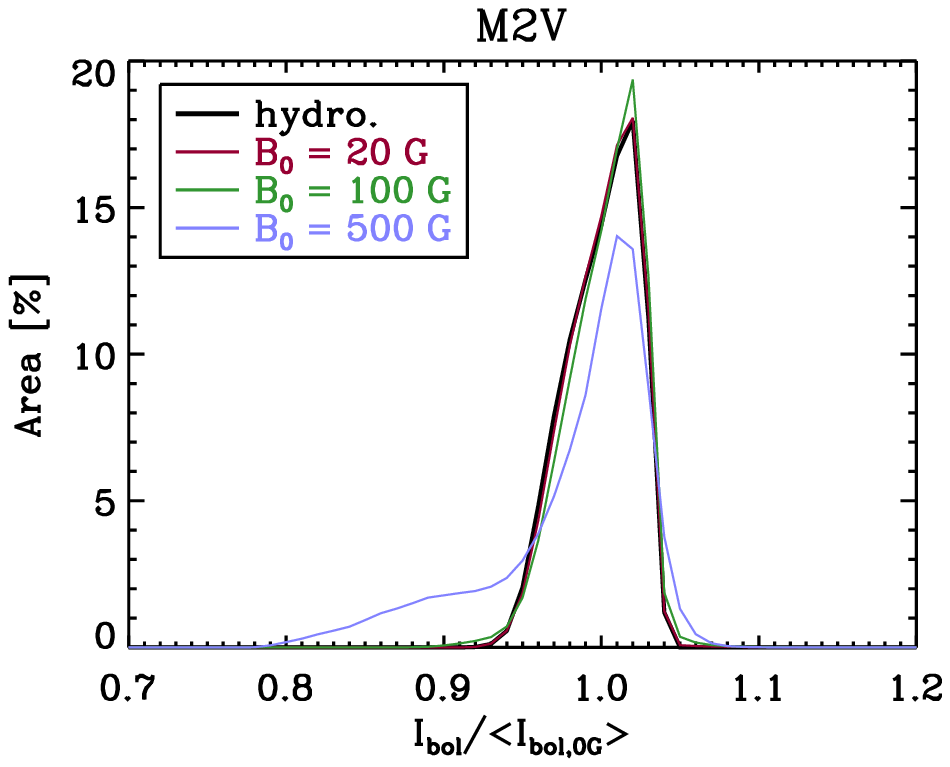}\\
\includegraphics[width=7.1cm]{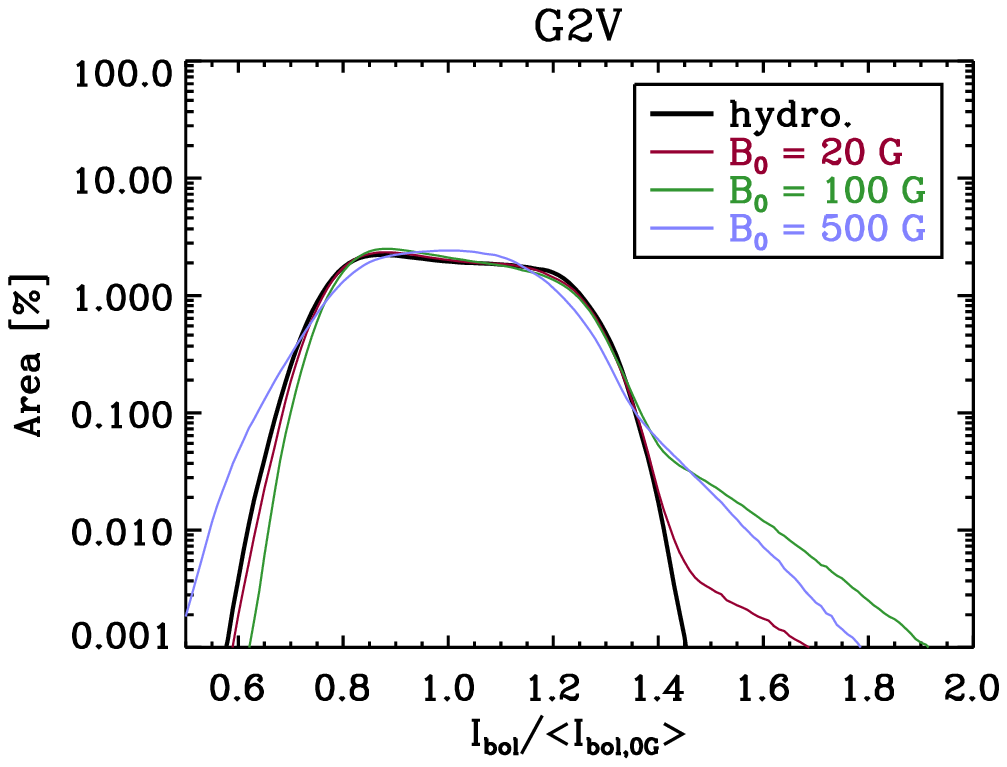}~\includegraphics[width=7.1cm]{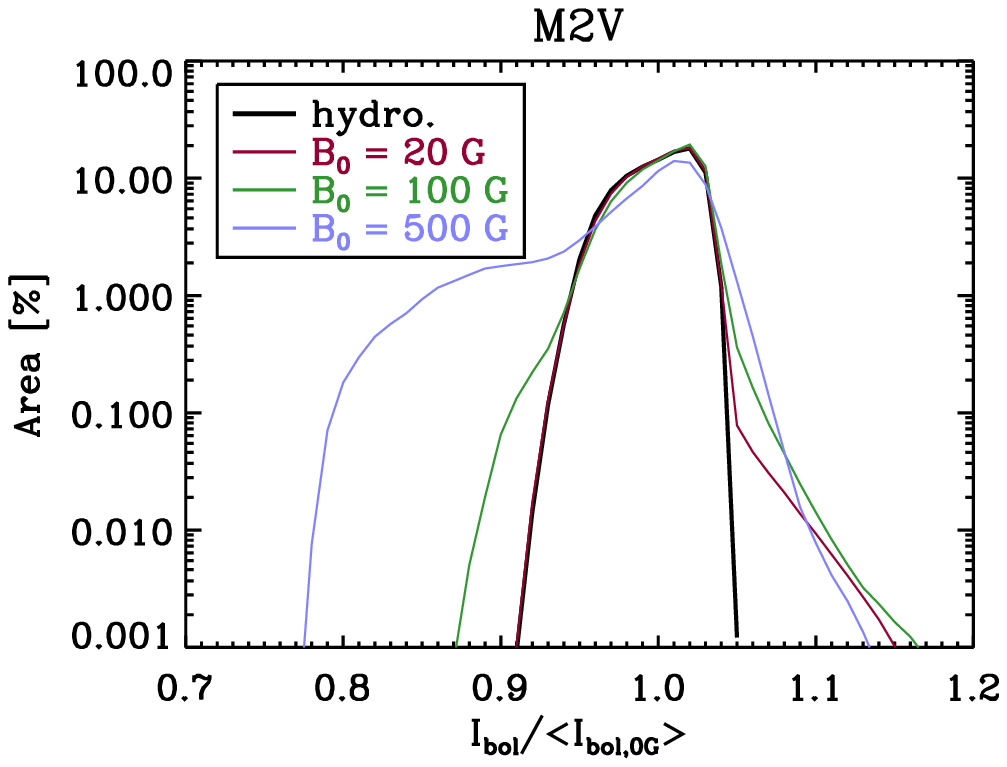}
\caption{Intensity histograms of the G2V ({\it left panels}) and M2V ({\it right panels}) simulations. In both representations ({\it upper panels}: linear; {\it lower panels}: logarithmic), the histogram bins are linearly equidistant with a bin size of 0.01.}\label{fig:I_hist}
\end{figure*}

In the F3V and G2V stars, the rms contrast is slightly reduced by the magnetic
field while minimum and maximum intensity are somewhat increased as
(moderately) bright magnetic regions replace the darkest regions in the
intergranular lanes. In the K stars, the contrast is increasing with
increasing $B_0$. Here, the bright magnetic regions are much more prominent
than in the hotter stars. This is also reflected in the maximum value of the
intensity, which is strongly enhanced in the magnetic K-star simulations to
values up to 20 standard deviation above the mean intensity. The reason for
this particular brightness of the bright magnetic regions is the weak
dependence of the opacity on temperature between 4000 and 5000\,K, which
renders the location of the photospheric transition more sensitive to density
than to temperature. Consequently, the partial evacuation in the magnetic flux
concentrations shifts the optical surface to particularly high temperatures in
these stars. A similar effect has been observed in vertical vortices in the
non-magnetic K-star simulations (see Paper~II). From K5V to M2V, the bright
magnetic regions become less bright (and less frequent) with decreasing
effective temperature along our model sequence and the maximum intensity
decreases (in terms of standard deviations above the mean).\par

In the K-star and M-star simulations, the global intensity minima decrease
from $B_0=20\,\mathrm{G}$ to $B_0=500\,\mathrm{G}$, owing to the formation of
dark magnetic regions. These dark regions become more pronounced with lower
effective temperature as the depth of the optical surface depression becomes
smaller compared to the horizontal size of the flux concentrations (see
Paper~III). Dark magnetic structures become dominating in the M-star
simulations and significantly reduce the mean disc-centre intensity in the
500\,G runs with respect to the non-magnetic runs of these stars.\par
\begin{figure*}
\centering
\includegraphics[width=7.1cm]{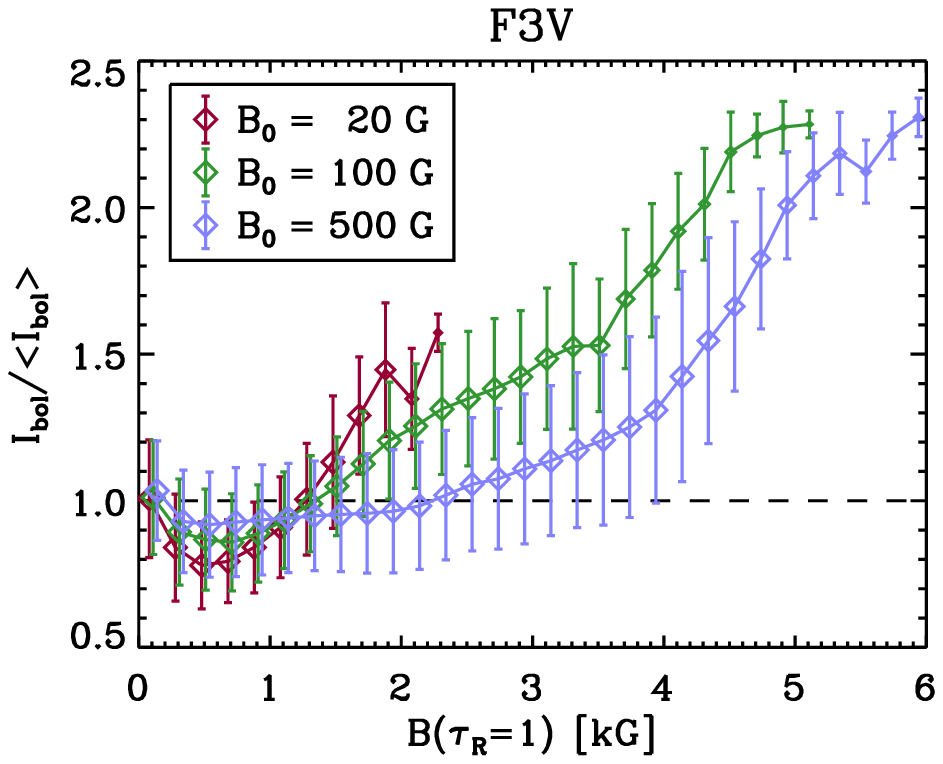}~\includegraphics[width=7.1cm]{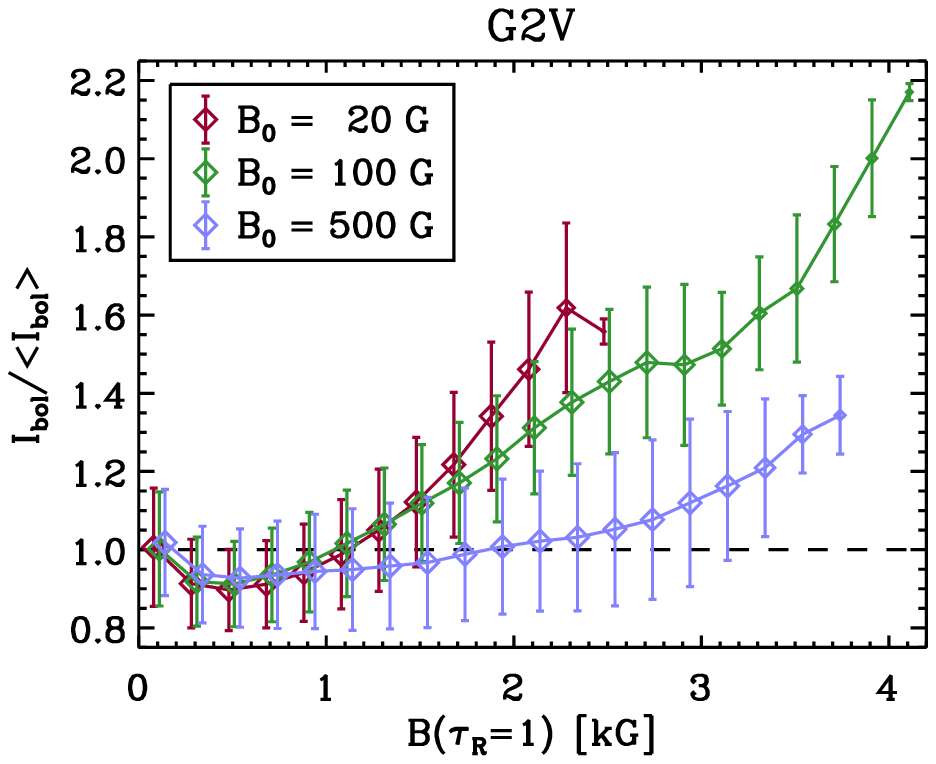}\\
\includegraphics[width=7.1cm]{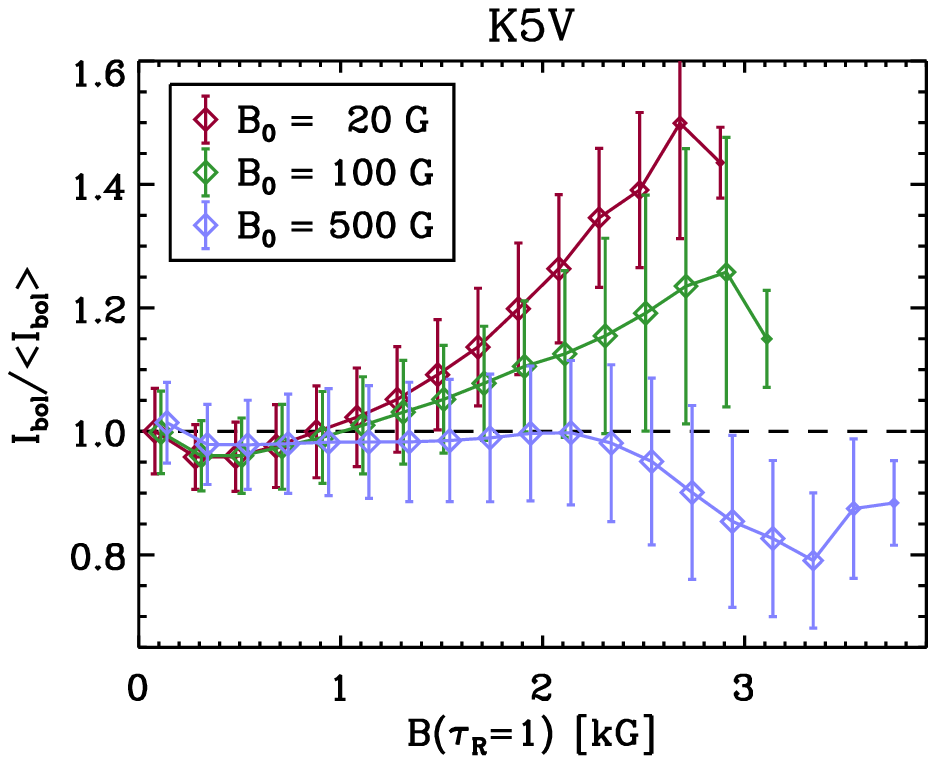}~\includegraphics[width=7.1cm]{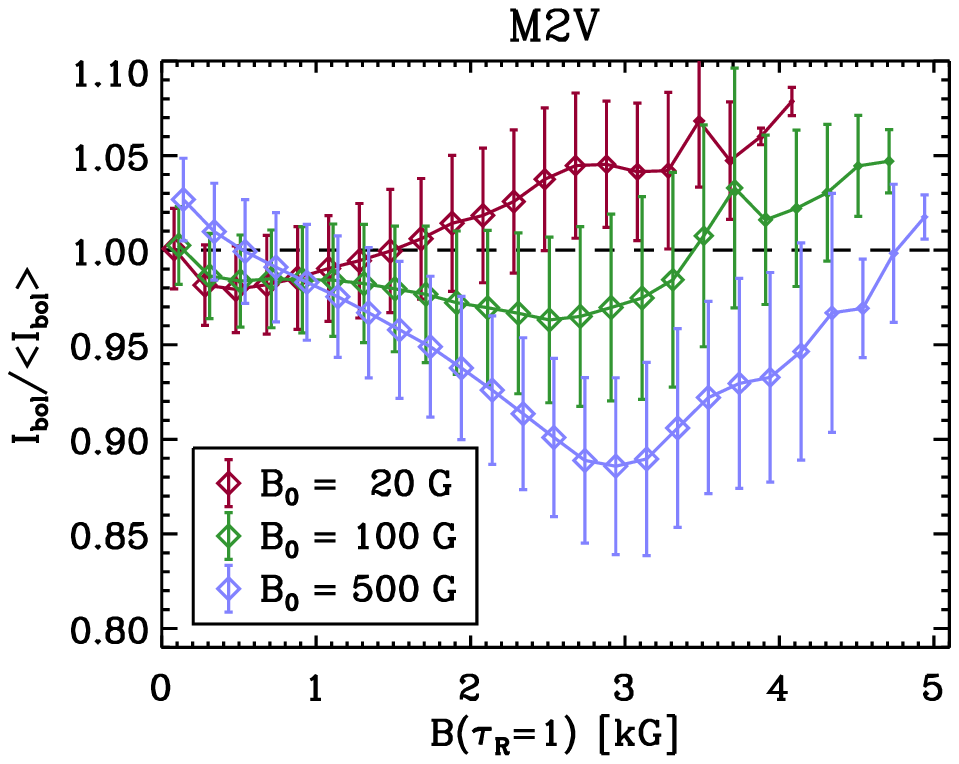}\\
\caption{Relation between the magnetic field strength $B(\tau_{\mathrm{R}}=1)$
  at the optical surface and the vertical bolometric intensity
  $I_{\mathrm{bol}}(\mu=1)$ normalised with its mean value. The field is
  binned in steps of 200\,G. Error bars give the 1-$\sigma$ scatter within the
  bins. For bins with fewer than 1000
  pixels ($\sim 0.06$\% of the total area considered), the symbol size is scaled
  with the logarithm of the number of pixels.}\label{fig:B_I}
\end{figure*}

Figure~\ref{fig:I_hist} shows histograms of the vertically emerging intensity
(normalised to the mean) for all four runs of the G2V and the M2V stars. The
bright magnetic features produce high-intensity tails in the histograms in all
magnetic runs of both stars. In the solar case, these bright features have the
highest area fraction and intensity in the 100\,G run and are considerably
weaker at $B_0=20\,\mathrm{G}$ and $B_0=500\,\mathrm{G}$. The dark magnetic
structures are responsible for low-intensity tails in the histograms of the
M2V-star simulations with $B_0=100\,\mathrm{G}$ and $B_0=500\,\mathrm{G}$ as
well in the 500\,G run of the G2V star. As pointed out in Paper~III, the
appearance of bright and dark magnetic structures does not only depend on
spectral type, but also on the amount of magnetic flux available (and hence on
$B_0$). Compared to the overall spread of the intensity distribution the dark
structures are much more prominent in the M2V star simulations than in the G2V
star (cf. Table~\ref{tab:I}).\par
%
\begin{figure*}
\centering
\includegraphics[width=7.1cm]{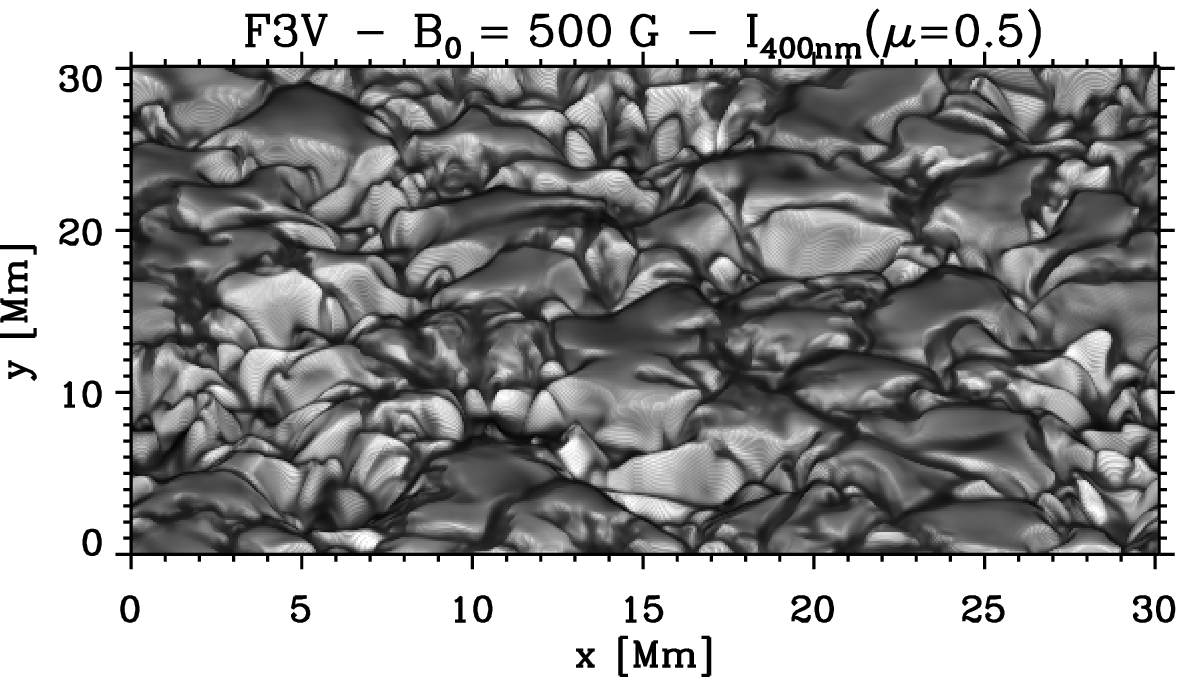}~\includegraphics[width=7.1cm]{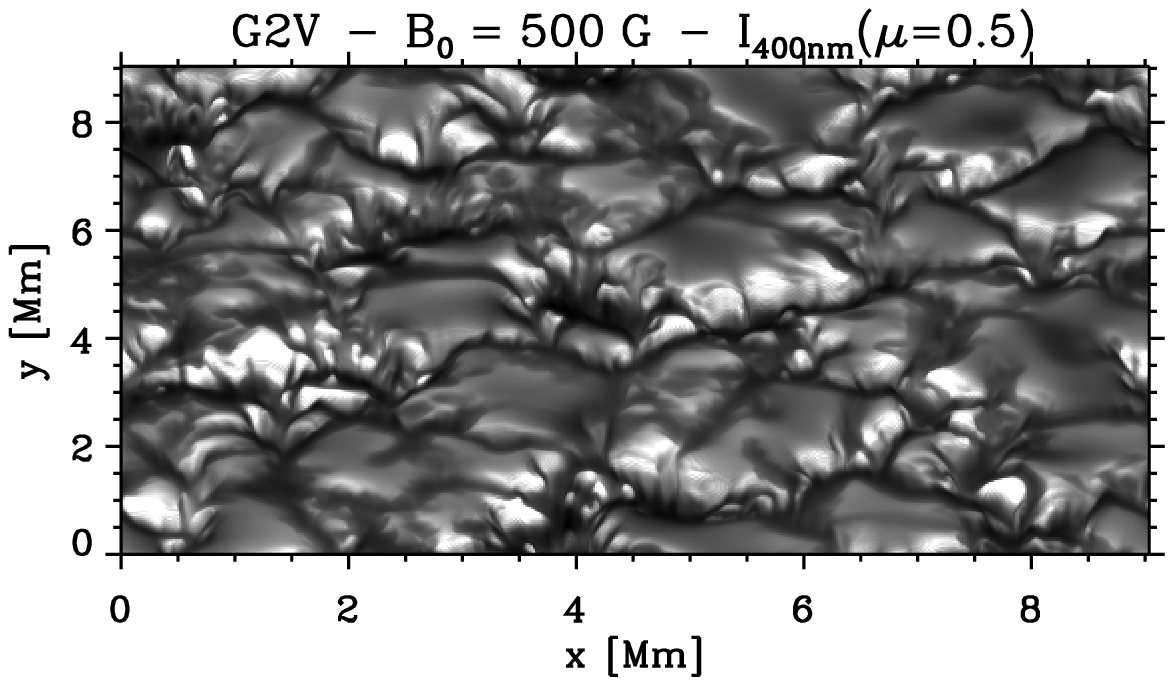}\\[5mm]
\includegraphics[width=7.1cm]{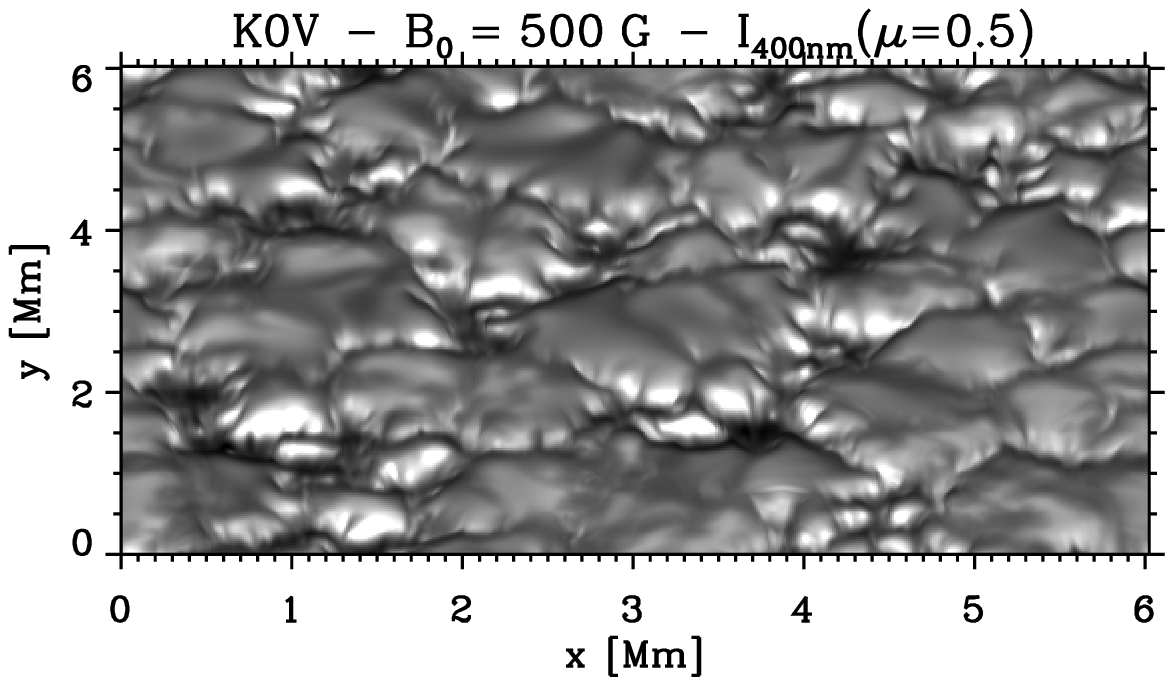}~\includegraphics[width=7.1cm]{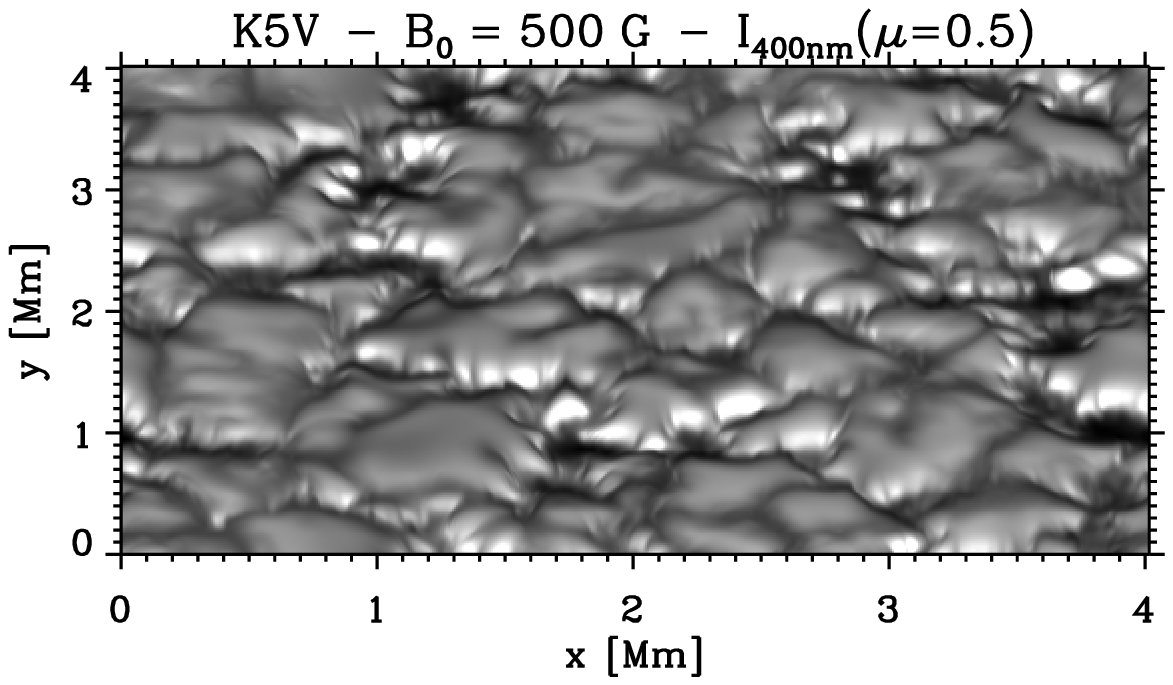} \\[5mm]
\includegraphics[width=7.1cm]{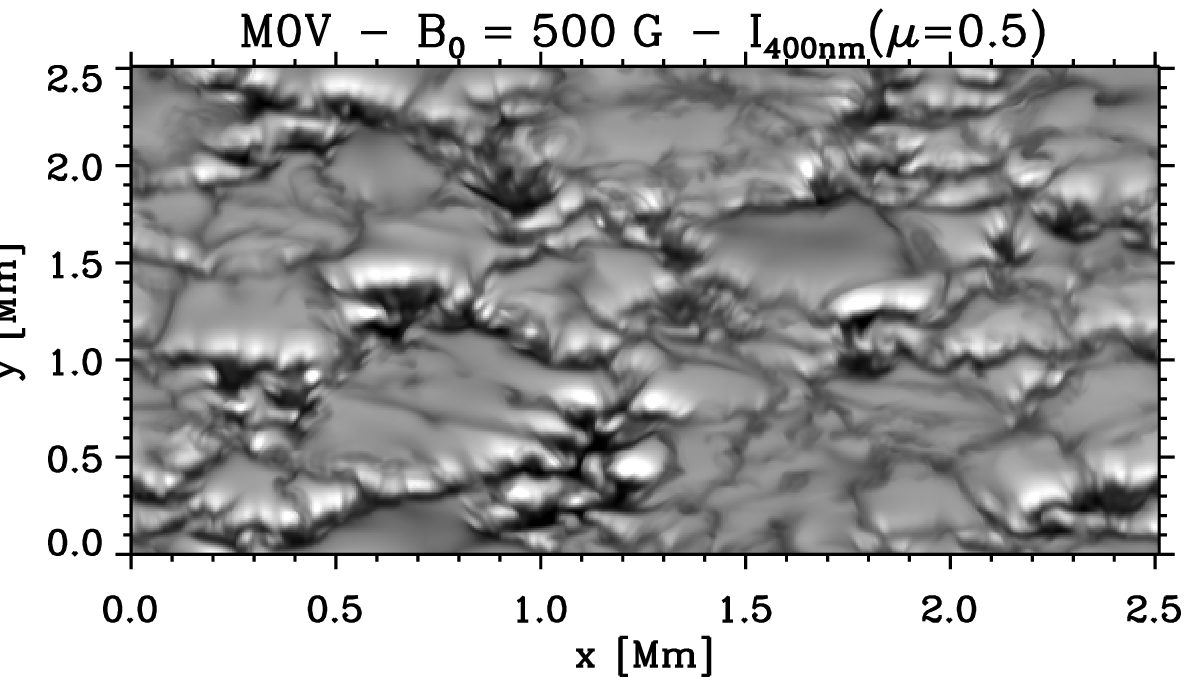}~\includegraphics[width=7.1cm]{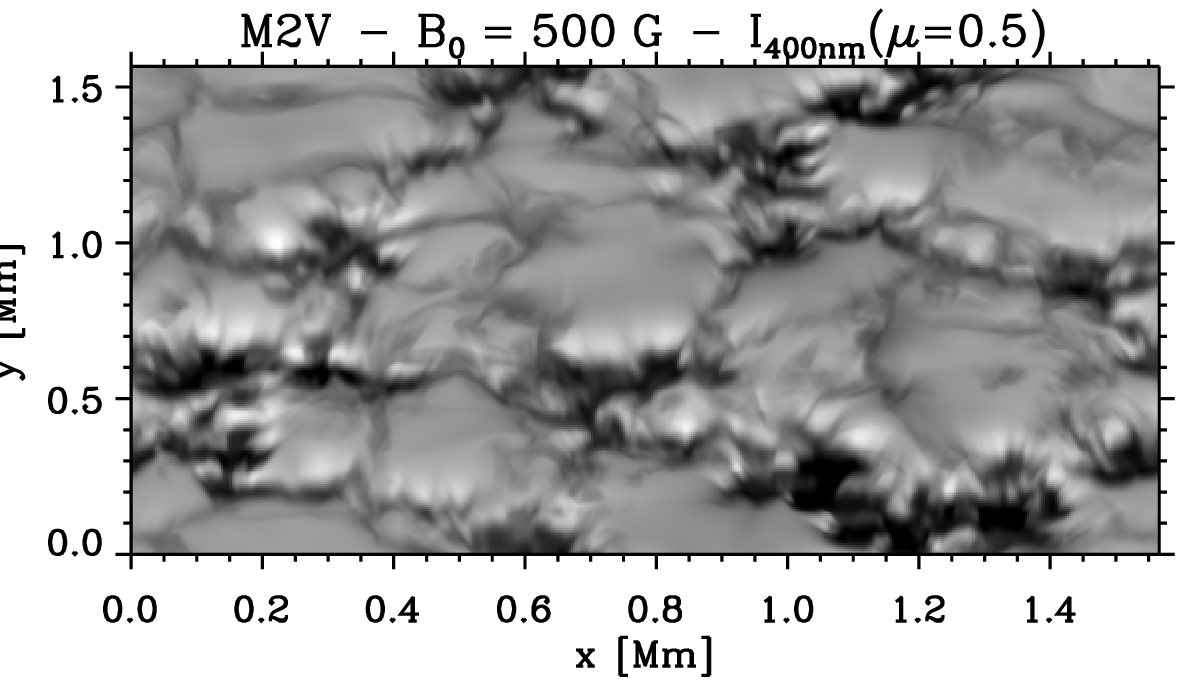}\\[5mm]
\caption{Continuum intensity in a passband between 400 and 410\,nm of the six simulations with $B_0=500\,\mathrm{G}$ viewed at an
  angle of $60^{\circ}$ with respect to the vertical
  (i.\,e. $\mu=0.5$).}\label{fig:inclview}
\end{figure*}
%
%
Figure~\ref{fig:B_I} shows a binned scatter plot of the vertical bolometric
intensity (normalised to its mean value) as a function of the magnetic field at
the optical surface for all magnetic runs of the F3V, G2V, K5V, and M2V
stars. In all runs (with the notable exception of the 500\,G run of the M2V
star), the mean intensity of the first bin
($B(\tau_{\mathrm{R}}=1)<200\,\mathrm{G}$), is close to the mean intensity of
the whole box, while the intensity of regions with a surface magnetic field
strength between 200 and 1000\,G on average falls somewhat below this
value. This has purely kinematic reasons: the magnetic flux is concentrated in
the dark intergranular lanes, which stay dark as long as the field is too weak
to considerably influence the thermodynamical structure and the flow
patterns. At a higher field strength ($B(\tau_{\mathrm{R}}=1) >
1\,\mathrm{kG}$), many of the curves have a positive slope, indicating that
regions of high magnetic field strength have, on average, an increasing
intensity with increasing field strength. This is the case for all runs for
the F3V and G2V stars. In these two models, the 20\,G runs show the highest
intensities at a given field strength. The intensity of highly magnetised
regions in the 500\,G runs is somewhat lower than in the 20\,G and 100\,G runs and the
scatter of intensity values (indicated by the error bars, which give the
1-$\sigma$ scatter) within each bin increases somewhat. This can be attributed to the
formation of a few dark magnetic structures among the still dominant component
of bright structures. In the K5V simulation with $B_0=500\,\mathrm{G}$ as well
as in the 100\,G and 500\,G runs of the M2V simulations, the slope of the
binned scatter plots stays negative in the kG regime indicating that dark
regions are dominant in these simulation runs. Only at very high field
strengths of $B(\tau_{\mathrm{R}}=0)> 3\,\mathrm{kG}$ does the average intensity
tend to increase with increasing field strength in these models. This is,
however, produced by very few surface elements, which correspond to
localised bright points with very small area coverage.\par

\subsection{Centre-to-limb variation of intensity and contrast}\label{sec:clv}
\begin{figure*}
\centering
  \includegraphics[width=7.1cm]{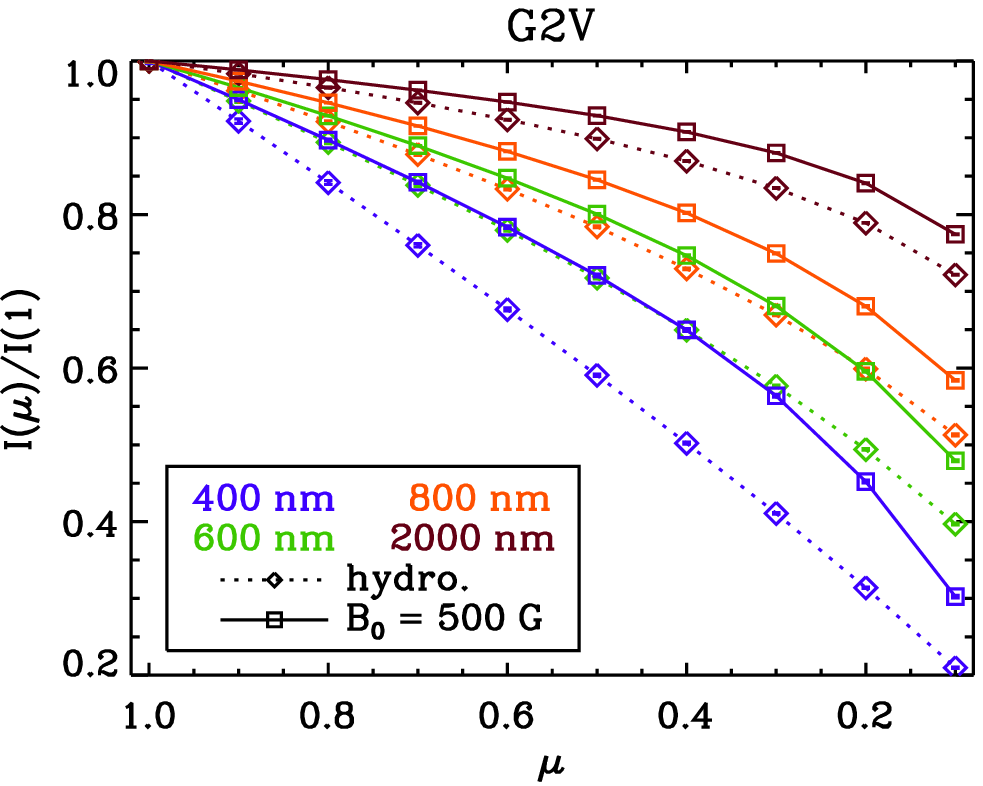}~~\includegraphics[width=7.1cm]{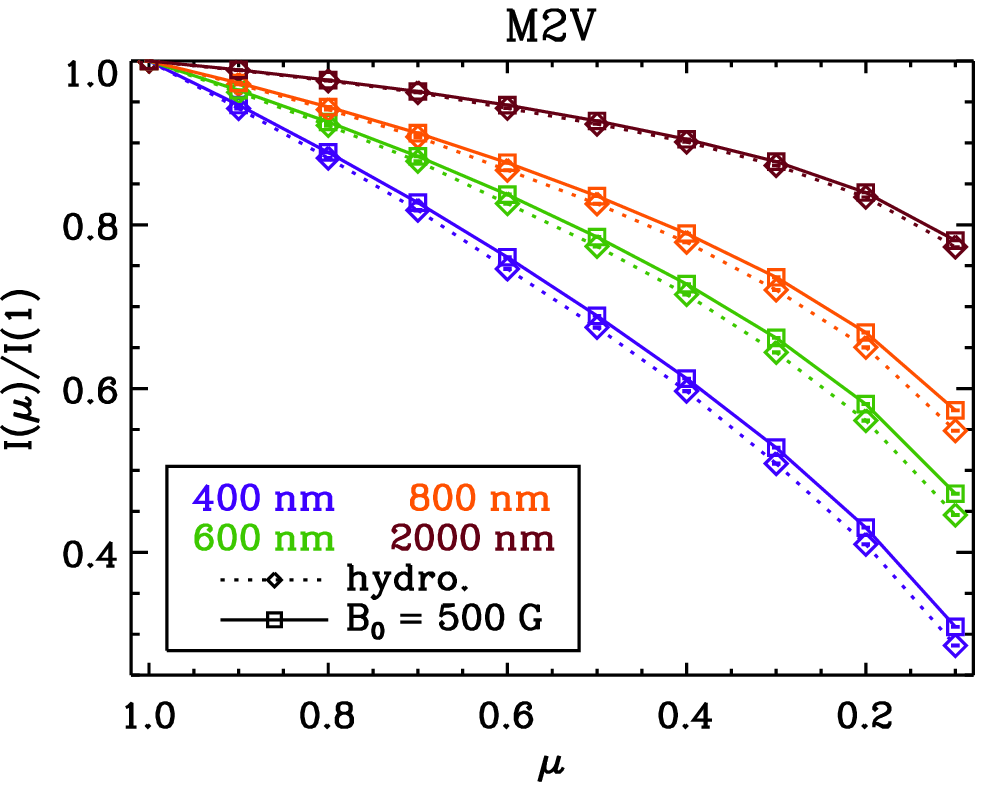}\\
\caption{Limb darkening for the G2V ({\it left}) and M2V ({\it right}) simulations without magnetic field ({\it diamonds and dotted curves}) and with $B_0=500\,G$ ({\it squares and solid curves}) in narrow passbands near 400, 600, 800, and 2000\,nm. The temporal scatter is considerably smaller than the symbol size.}\label{fig:LD1}
\end{figure*}
Figure~\ref{fig:inclview} shows maps of the continuum intensity near 400\,nm
of the 500\,G runs at an inclined viewing angle of $\mu = \cos \vartheta =
0.5$. The snapshots are the same as in Fig.~1 of Paper~III (and for G2V and
M2V the same as in Fig.~\ref{fig:I_map} in this paper). The differences
between the different spectral types are mainly caused by the different depth
of the depressions in the optical surface caused by the magnetic flux
concentrations, but also by the different impact of the flux concentrations on
the local density and temperature structures. In the F3V star, the depressions
are deepest (in terms of pressure scale heights as well as in terms of
granule sizes, see Paper~III) and the evacuation in the upper part of the
magnetic flux concentration is strongest. Consequently, the intergranular
lanes and the centres of the magnetic flux concentrations are hidden from view
behind granules while the granules which are seen through the highly
evacuated upper parts of magnetic structures appear brightened. Moreover, the
granules show dark edges where the optical path through the cool layers above
the optical surface is particularly long (``limb darkening'' of single
granules). With decreasing effective temperature the corrugation of the
optical surface as well as the typical depth of the depressions caused by the
magnetic flux concentrations decrease (see Paper~III) and the appearance of
$I(\mu=0.5)$ changes. In the G-, K-, and M-star simulations, only a part of the
granules seen through flux concentrations are brightened significantly. These
small scale brightenings are observed on the Sun as ``faculae''
\citep[see][and references therein]{Keller04}. In the M stars, especially M2V,
the dark magnetic structures are still visible at $\mu=0.5$ and the faculae
are less bright (in terms of the standard deviation of the intensity) than in
the K- and G-type stars.\par
%
\begin{figure*}
\centering
\includegraphics[width=7.1cm]{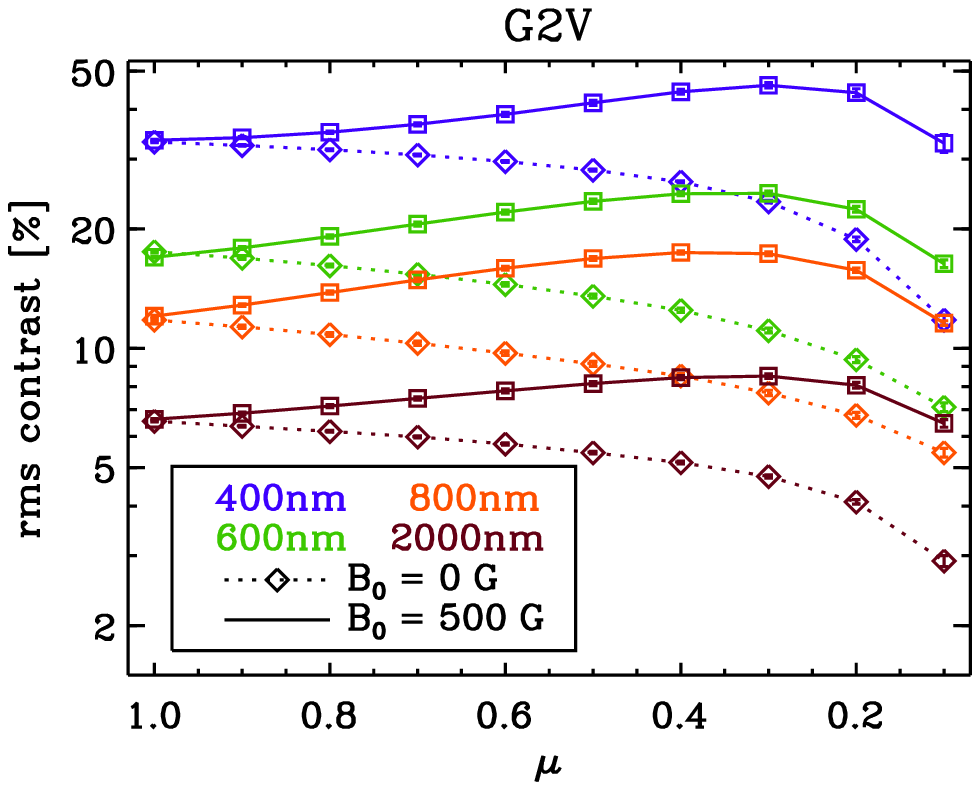}~~~~\includegraphics[width=7.1cm]{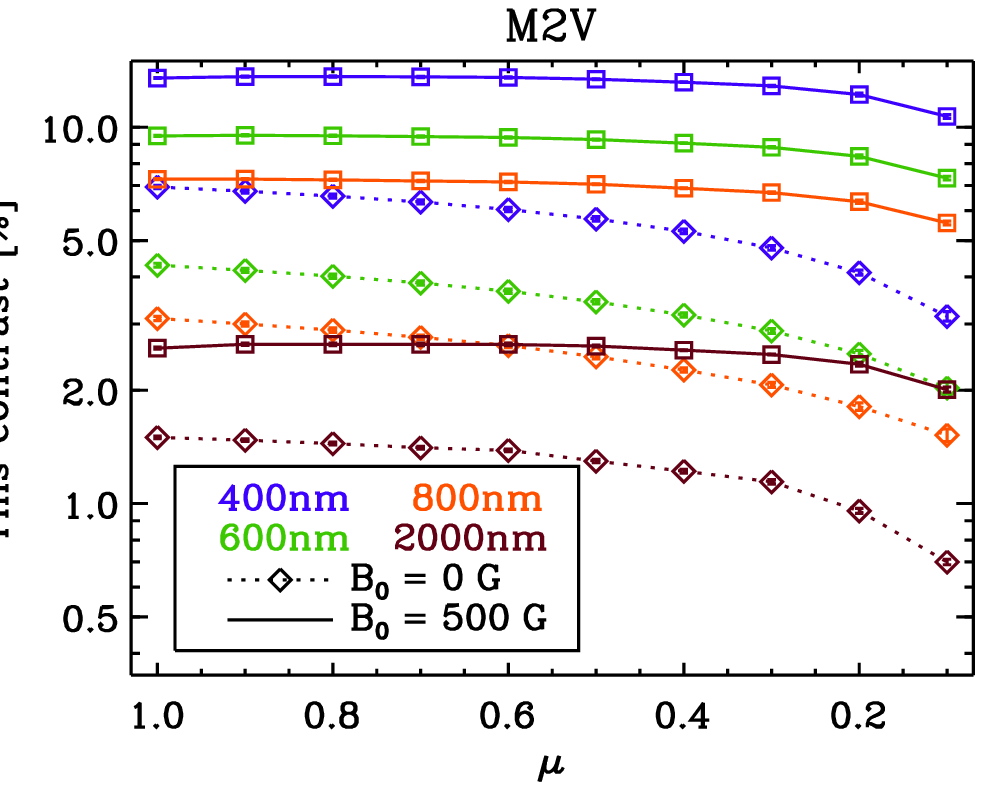}\\
\caption{Centre-to-limb variation of the rms intensity contrast for the G2V ({\it left}) and M2V ({\it right}) simulations without magnetic field ({\it diamonds and dotted curves}) and with $B_0=500\,G$ ({\it squares and solid curves}) in narrow passbands near 400, 600, 800, and 2000\,nm. The temporal scatter is considerably smaller than the symbol size. We note the logarithmic scale of the ordinates.}\label{fig:LD3}
\end{figure*}
%
Figure~\ref{fig:LD1} shows the centre-to-limb variation of the continuum
intensity in four narrow wavelength passbands for the G2V- and the M2V-star
simulations, comparing the non-magnetic runs (dotted curves) with the 500\,G
runs. In both cases, the limb darkening is reduced in the magnetic runs
compared to the non-magnetic runs, i.\,e. the magnetic field effectively
brightens the limb. This limb brightening is produced by the faculae
(cf. Fig.~\ref{fig:inclview}). The effect is stronger at higher effective temperature as the faculae
become more prominent and have a higher filling factor. In the G2V-star
simulation, the difference in limb darkening between the non-magnetic and the
500\,G runs is even larger than the difference between the non-magnetic runs
of different spectral types.\par
%
\begin{figure*}
\centering
\includegraphics[width=7.1cm]{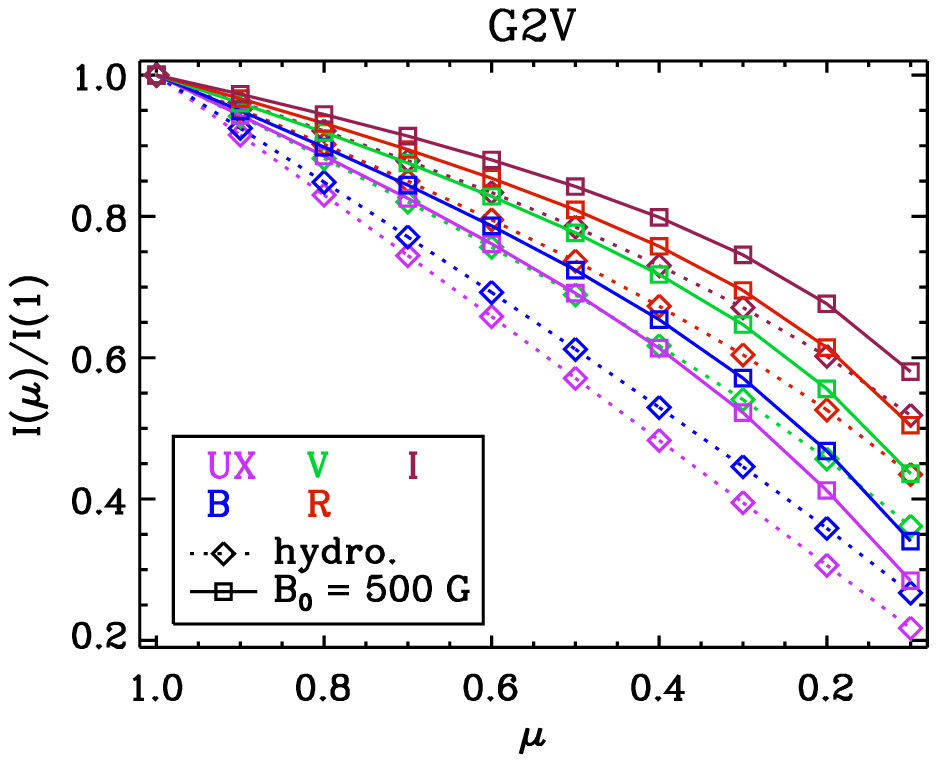}~\includegraphics[width=7.1cm]{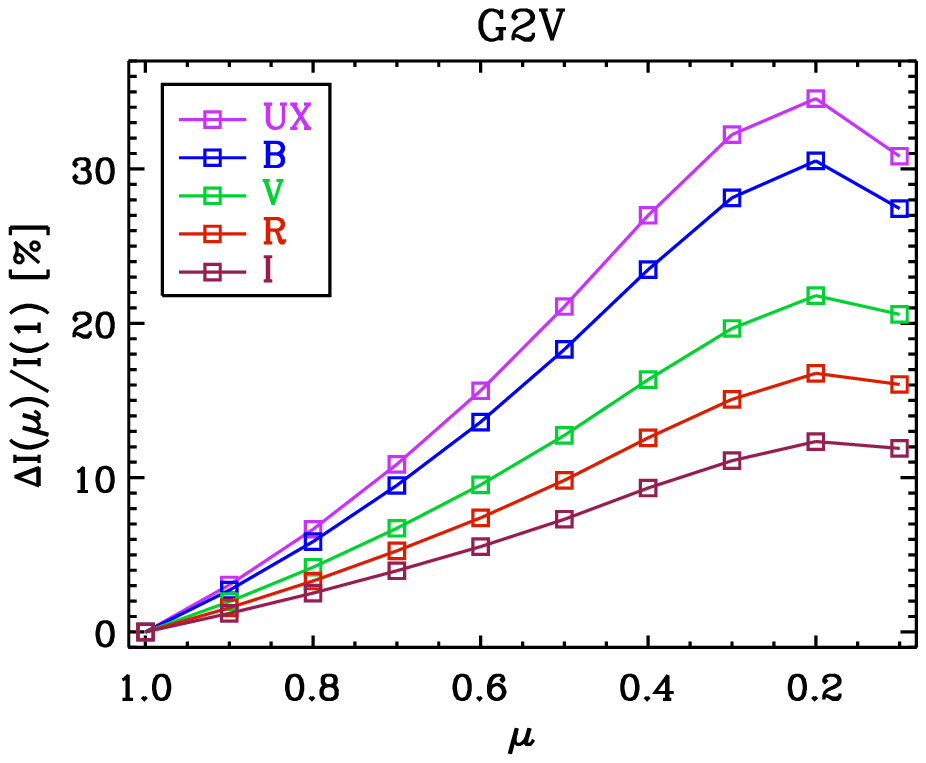}
\caption{Limb darkening in the Johnson UX, B, V, R, and I passband filters \citep{Johnson, Bessell90} in the G2V star. Analogous plots for the F3V and the M0V star are given in Fig.~\ref{fig:LD_app} in Appendix~\ref{app:figs}. {\it Left panel:} Limb darkening of the non-magnetic ({\it hydro.}) and 500\,G runs. {\it Right panel:} Relative difference between both runs.}\label{fig:LD_Johnson}
\end{figure*}
%
In Figure~\ref{fig:LD3}, the centre-to-limb variation of the relative
intensity fluctuations (``intensity contrast'') is plotted as a function of
$\mu=\cos\vartheta$ for the non-magnetic and 500\,G runs of the solar (G2V)
and M2V-star simulations. In the solar case, the intensity contrast at the
disc centre does not significantly differ between non-magnetic case and 500\,G
case (cf. Table~\ref{tab:I}). Near the limb the effect of the faculae leads to
a strongly enhanced intensity contrast: while in the non-magnetic case, the
contrast monotonically drops from the disc centre to the limb, it reaches a
maximum in the 500\,G runs around $\mu=0.3$ in all wavelength passbands
considered. In the M2V-star simulations, the magnetic field affects the
intensity contrast at all disc positions: the contrast is roughly twice (near
the limb: three times) as high in the magnetic run as in the non-magnetic
run. Unlike the solar case, the intensity contrast does not strongly increase
from the disc centre towards the limb in the 500\,G run as the faculae are
less pronounced.\par
%
%
\begin{figure*}
\centering
\includegraphics[width=8.5cm]{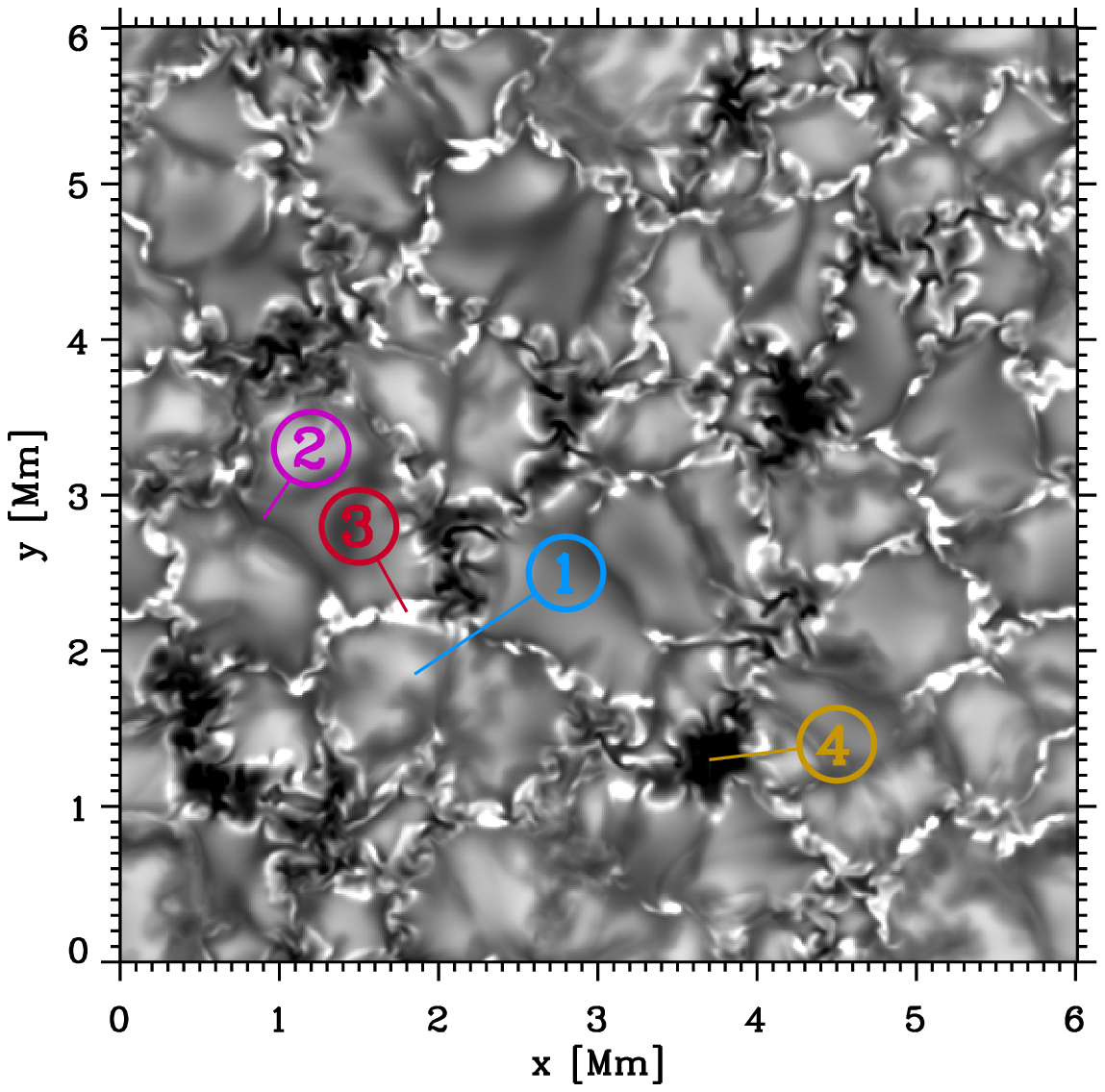}~\parbox[b]{5.7cm}{\includegraphics[width=5.7cm]{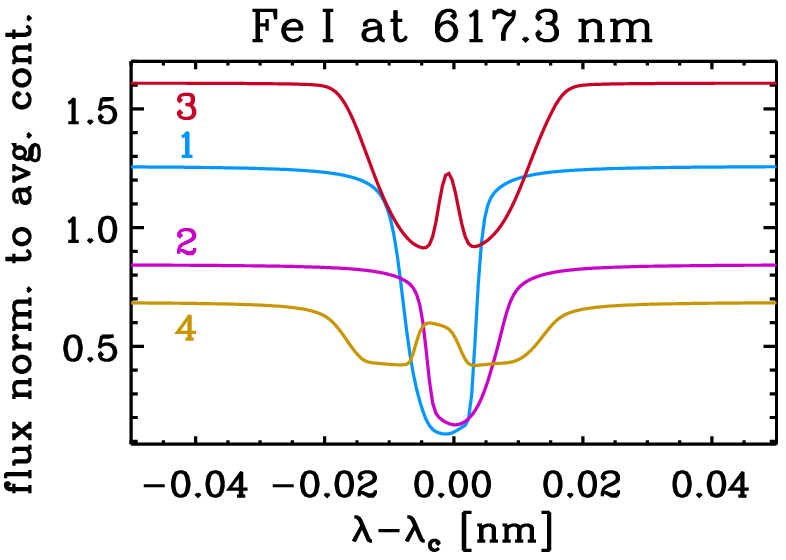}\\\includegraphics[width=5.7cm]{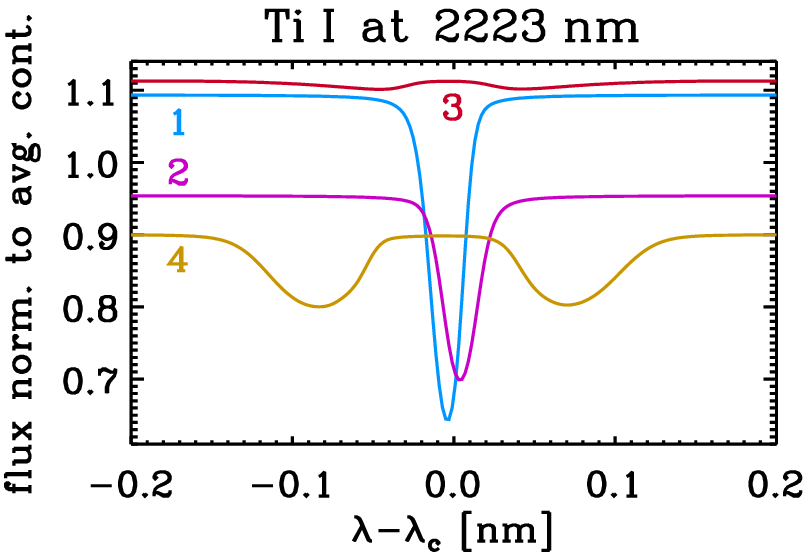}}
\caption{Local vertical spectral line profiles for the K0V simulation with $B_0=500\,\mathrm{G}$. {\it Left panel:} intensity image with four positions marked. {\it Right panels:} local profiles of the Fe\,\textsc{i} line at 617.3\,nm ({\it top}) and the Ti\,\textsc{i} line at 2223\,nm ({\it bottom}) at ($\mu=1$); the four different colours correspond to the four points marked in the left panel.}\label{fig:local_spectra}
\end{figure*}
%
%

Figure~\ref{fig:LD_Johnson} shows the limb darkening in the non-magnetic and
the 500\,G runs of the G2V star in the Johnson UX, B, V, R, and I passbands
(left panel) and the difference between the two runs in per cent (right
panel). The highest relative deviations are in the UX band near the limb,
where the magnetic run is more than 30\% brighter than the non-magnetic
run. For some real stars, limb darkening can be inferred from eclipses (in
binaries) or exoplanet transits. Possibly, the limb darkening can be used as
an indicator for small scale magnetic structures (faculae) on such stars even
in cases where the field is too small to be detected spectroscopically. In the same way as for
the continuum passbands (see Fig.~\ref{fig:LD1}), the magnetic limb
brightening becomes weaker with decreasing effective temperature along our
model sequence in the Johnson passbands (see Fig.~\ref{fig:LD_app} in
Appendix~\ref{app:figs}).

\section{Spectral line profiles}
\subsection{Spectral line profiles at the disc centre}\label{sec:local}
In Paper~II, we analysed the effects of the convective velocity field and its
spatial correlation with the thermodynamical quantities on a set of three
spectral lines (Fe\,\textsc{i} at 616.5\,nm, Fe\,\textsc{i} at 617.3\,nm, and
Ti\,\textsc{i} at 2223.3\,nm). Here, we investigate the impact of the local
magnetic field structure and its correlation with thermodynamical quantities
and with the convective flows on these lines. In the presence of a magnetic
field, all three lines are split into their Zeeman components. The effective
Land\'e factors are quite different: while the Fe\,\textsc{i} line at
617.3\,nm was chosen for its large effective Land\'e factor of
$g_{\mathrm{eff}}=2.5$, the Fe\,\textsc{i} line at 616.5\,nm is rather
insensitive to the magnetic field ($g_{\mathrm{eff}}=0.69$). The third line,
Ti\,\textsc{i} at 2223.3\,nm, has an effective Land\'e factor of
$g_{\mathrm{eff}}=1.66$. However, as it is in the infrared, the broadening or
splitting due to the Zeeman effect is expected to be larger in comparison to
the line width than in the optical Fe\,\textsc{i} lines.\par
%
%
Figure~\ref{fig:local_spectra} shows a snapshot of the K0V-star simulation
with $B_0=500\,\mathrm{G}$. Four pixels are marked for which vertical local
spectral line profiles of the Ti\,\textsc{i} line and the Fe\,\textsc{i} line
at 617.3\,nm are plotted in the two right panels. Point 1 is located in an upflow region with
negligible magnetic field strength. The corresponding line profiles are
shifted somewhat to the blue. Especially for the Fe\,\textsc{i} line at 617.3\,nm, this
shift affects the line wings more strongly than the core, resulting in a
strong asymmetry. As the wings of these photospheric lines form very close to
the surface, while the line core forms somewhat higher in the stellar
atmosphere, this asymmetry reflects the deceleration of the (overshooting)
upflow as discussed in Paper~II. Point 2 is located in a downflow with low
field strength. The corresponding line profiles are shifted to the red, with
an asymmetry (especially in the Fe\,\textsc{i} line at 617.3\,nm) indicating an
acceleration along the downflow. We note that the continuum flux is lower in
Point 2 and that the depth of the line profiles is smaller. In the average
spectrum, Point 2 consequently has a lower weight than Point 1 for these lines.\par 
%
\begin{figure*}
\centering
\includegraphics[width=7.1cm]{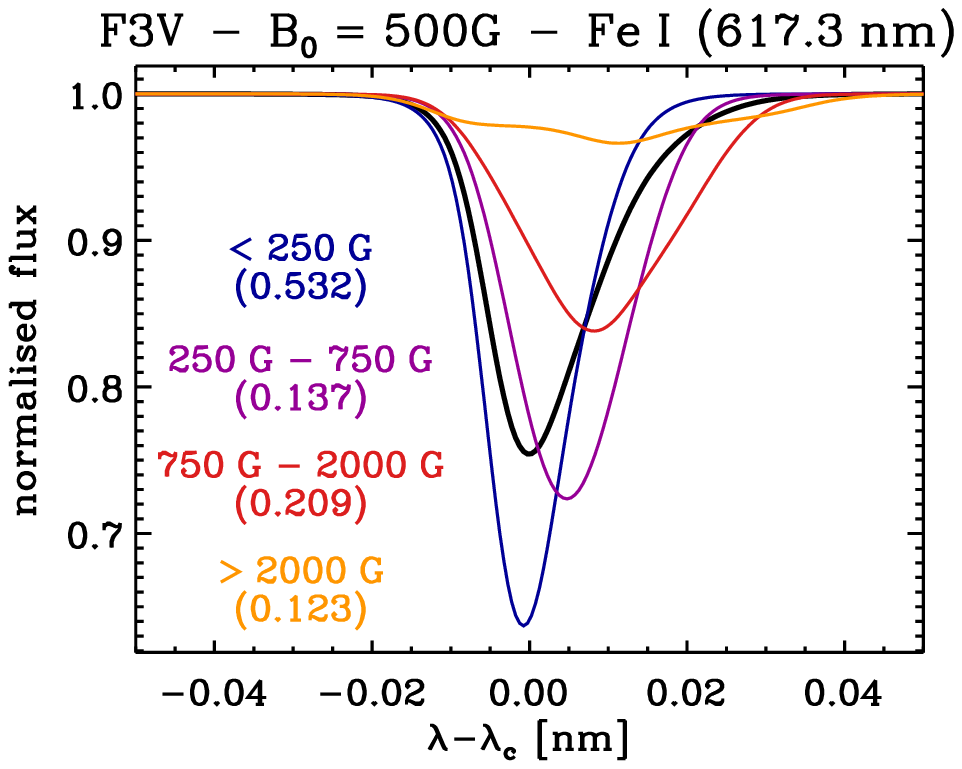}~\includegraphics[width=7.1cm]{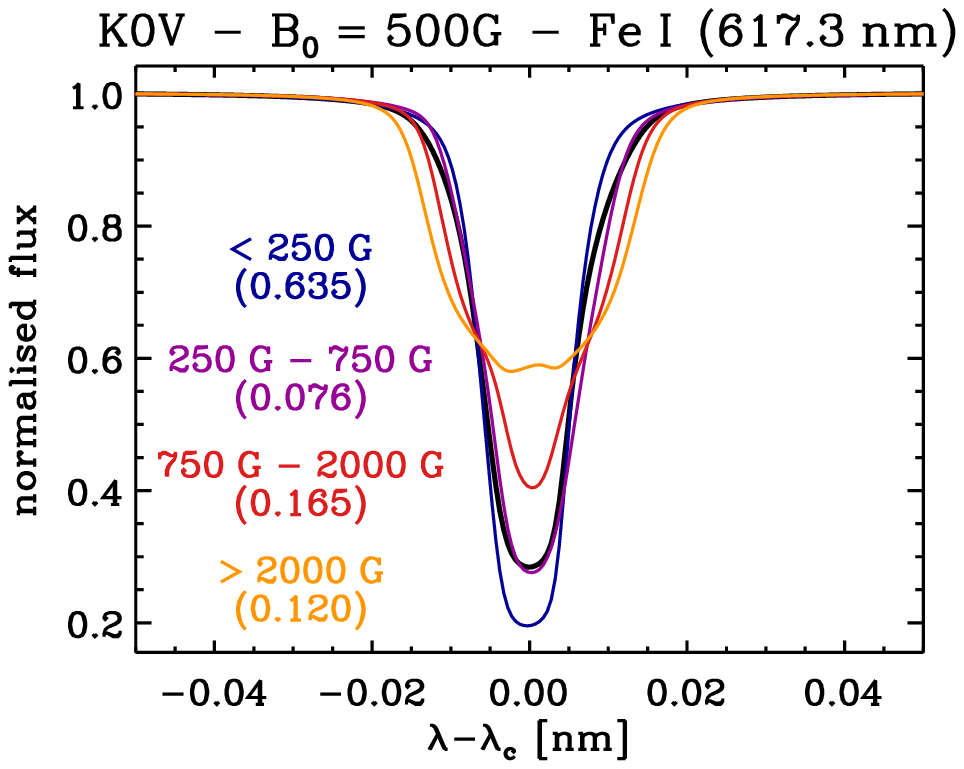}\\
\includegraphics[width=7.1cm]{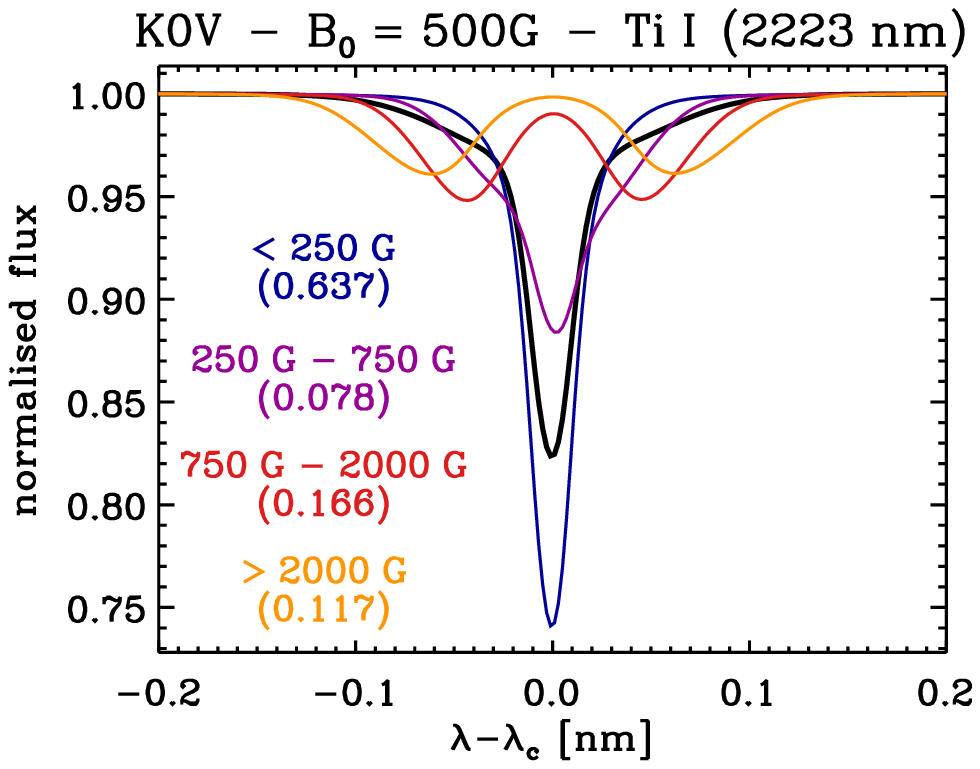}~\includegraphics[width=7.1cm]{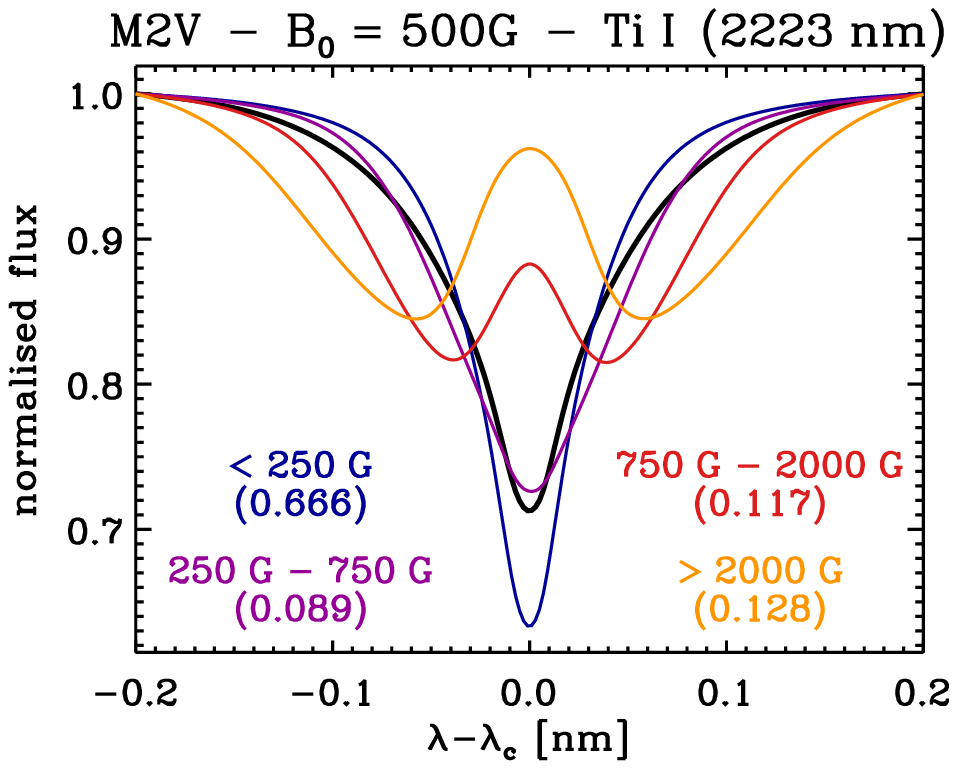}
\caption{Decomposition of four line profiles (at $\mu=1$) into differently
  magnetised components. The black curve shows the average line profile, while
  the coloured curves show the line averaged over areas where
  $B(\tau_{\mathrm{R}}=1)$ is in the specified ranges. Each line is normalised
  to its own continuum. The numbers in parentheses give the weight of the
  components in the composite spectrum (i.\,e. average continuum flux times
  area fraction).} \label{fig:line_contribution}
\end{figure*}

The other two points in Fig.~\ref{fig:local_spectra} are situated in
magnetised regions: Points~3 in a bright (hot) magnetic structure and Point~4
in the centre of a dark (cool) magnetic structure. The corresponding spectral
line profiles are split into two components by virtue of the Zeeman
effect. The components are asymmetric owing to the vertical gradient of the
magnetic field: as the field rapidly decreases with height in the photosphere
(see Paper~III), the cores of the Zeeman components are less strongly shifted
away from the line centre than their wings. In comparison to Point~1, the
equivalent width (EW) of the Fe\,\textsc{i} line at 617.3\,nm (upper
right panel of Fig.~\ref{fig:local_spectra}) is larger in Point~3 and smaller
in Point~4, caused by the (moderate) temperature sensitivity of the
corresponding line opacity. As the continuum is also higher in Point~3 than in
Point~4, the weight of the local line profile from Point~3 will be higher in
the spatially averaged line profile than that of Point~4.\par In contrast to
the Fe\,\textsc{i} line at 617.3\,nm, the Ti\,\textsc{i} line (lower
right panel in Fig.~\ref{fig:local_spectra}) has a strongly reduced EW in the
bright magnetic structure (Point~3) with respect to Points~1 and 2. The
Ti\,\textsc{i} line is very sensitive to temperature around 4000 K, owing to
ionisation and to depopulation of the lower level of the
transition. Therefore, the line opacity is considerably reduced in hot
magnetic structures of this star. This effect will be referred to as ``line
weakening'' in the following discussion. The same phenomenon also
  exists in the Fe\,\textsc{i} lines but at a higher temperature (around
  6000\,K), which is more relevant for the G2V- and F3V-star simulations. In the dark magnetic structure (Point~4), the EW of the
Ti~\textsc{i} line is somewhat larger than in Points~1 and 2. In summary, dark
(cool) and bright (hot) magnetic structures can produce crucially different
local spectral line profiles. As shown in Sect.~\ref{sec:vert_int}, the
relative importance of dark and bright magnetic structures depends on the
average field strength and the stellar parameters.\par
%
%
Figure~\ref{fig:line_contribution} shows horizontally averaged profiles of the
Fe\,\textsc{i} line at 617.3\,nm for the 500\,G runs of the F3V- and the
K0V-star simulations and of the Ti\,\textsc{i} line for the 500\,G runs of the
K0V- and the M2V-star simulations. The thick black curves represent the line
profiles averaged over the entire area of the simulation box, while the four
coloured curves are the profiles averaged over surface components with
different magnetic field ranges at the optical surface. The numbers in
parentheses give the weight of the individual components in the average line
profile. The profile of the Fe\,\textsc{i} line at 617.3\,nm has a strongly extended red
wing in the F3V simulations. As the decomposition into distinct surface
components illustrates, this red wing is produced mainly by the flow
velocities in the area components with intermediate field strengths of 250 to
2000\,G: these components mainly represent short-lived, dynamic magnetic flux
concentrations in a convective collapse phase in this star (see
Paper~III). This results in an (on average) increased downflow velocity in the
optical depth range where this spectral line is forming. Additionally, the
Zeeman effect broadens the line profiles emerging from these surface
components, shifting its red wing even further towards longer wavelengths. The
regions with high magnetic field strengths of $>2\,\mathrm{kG}$ do not
contribute much to the average spectrum because of line weakening, which
affects the Fe\,\textsc{i} lines in the G2V and F3V atmospheres (while it does
not play a great role for K0V, cf. Fig.~\ref{fig:local_spectra}).\par 
%
\begin{figure*}
\centering
\includegraphics[width=4.4cm]{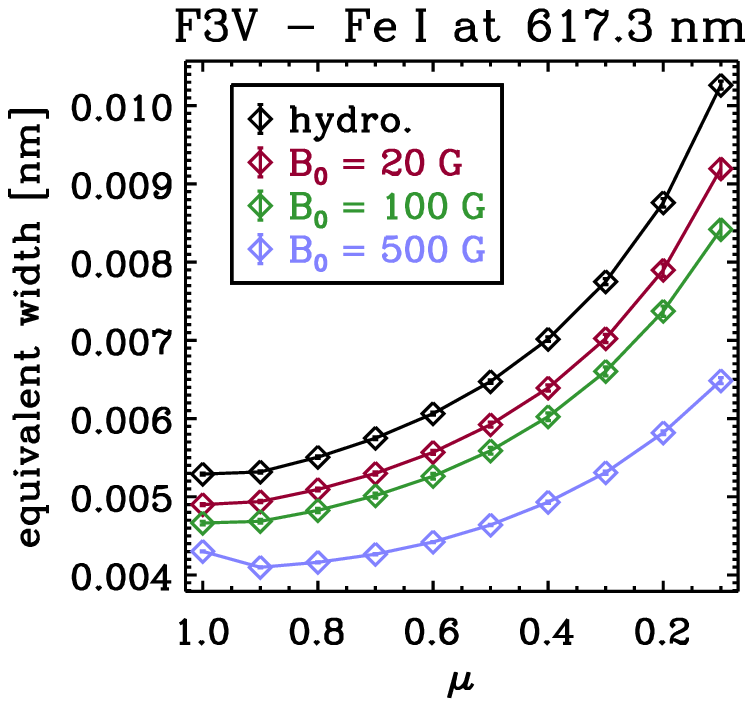}\includegraphics[width=4.4cm]{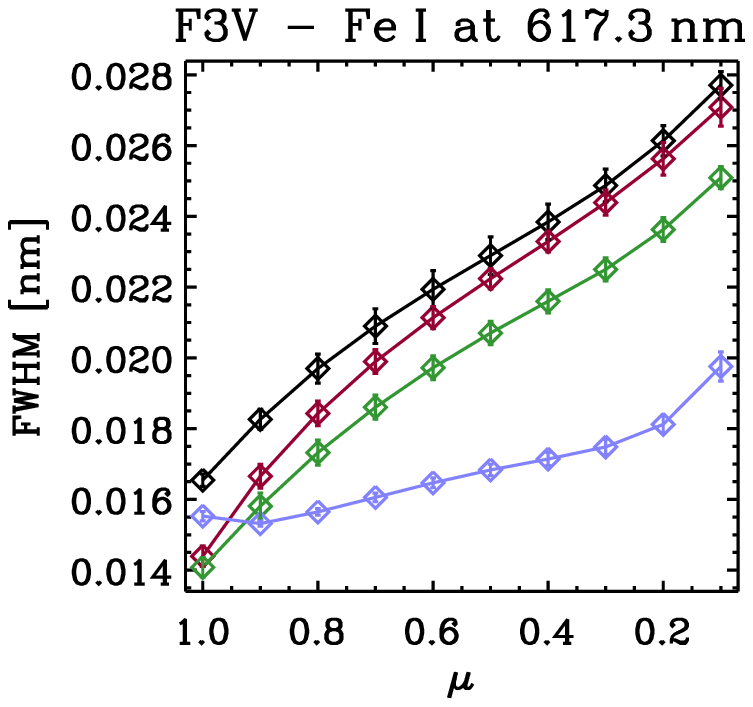}\includegraphics[width=4.4cm]{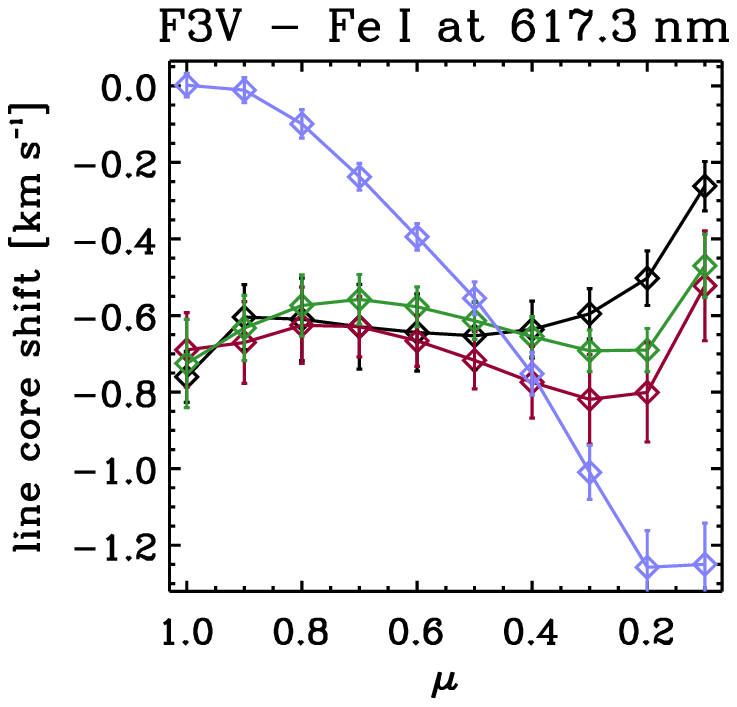}\includegraphics[width=4.4cm]{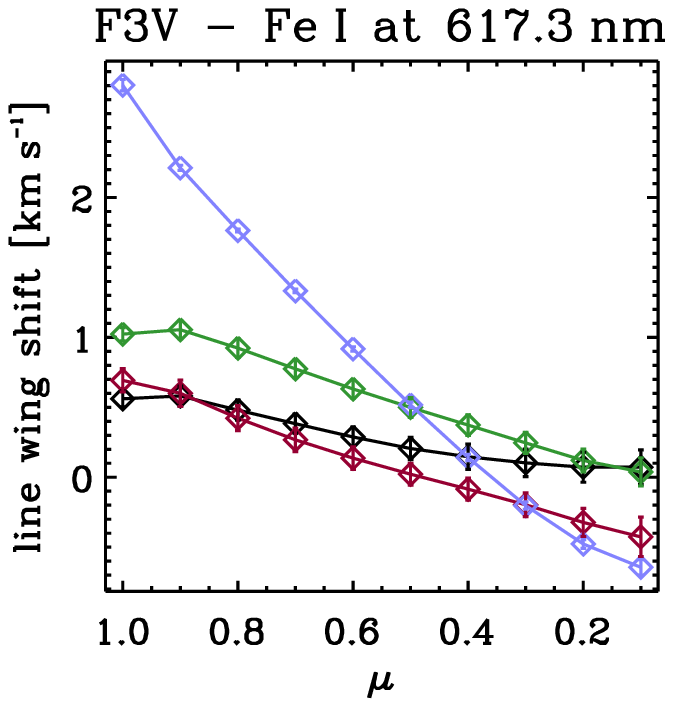}
\caption{Centre-to-limb variation of the equivalent width, FWHM, and the effective Doppler shifts of the line core and wings ({\it from left to right}) of the Fe\,\textsc{i} line at 617.3\,nm in all four simulation runs of the F3V star. The colour code is the same for all panels, see legend in the left panel.}\label{fig:lclv_F3}
\end{figure*}

In the K0V simulation, the flow velocities are much lower and the magnetic
flux concentrations are less dynamic. Consequently, the Doppler shift of the
line components is weaker. As the temperature is lower, less line weakening
occurs in the Fe\,\textsc{i} line at 617.3\,nm and the substantially broadened line profile
emerging from the strongly magnetised area components produces extended wings
in the averaged (disc-centre) line profile. For the Ti\,\textsc{i} line, these
wings are even more pronounced as the Zeeman splitting is stronger in the
infrared (with respect to the line width). Owing to line weakening, however,
the equivalent width of the line profiles stemming from the magnetised area is
somewhat reduced. In the M2V simulation, line weakening does not play a role
for the lines considered. The line width is larger here for all surface
components owing to the higher photospheric pressure, rendering the extended
wings more difficult to see in the pressure-broadened line profile.
\subsection{Centre-to-limb-variation of spectral line profiles}\label{sec:lclv} 
%
\begin{figure*}
\centering
\includegraphics[width=4.1cm]{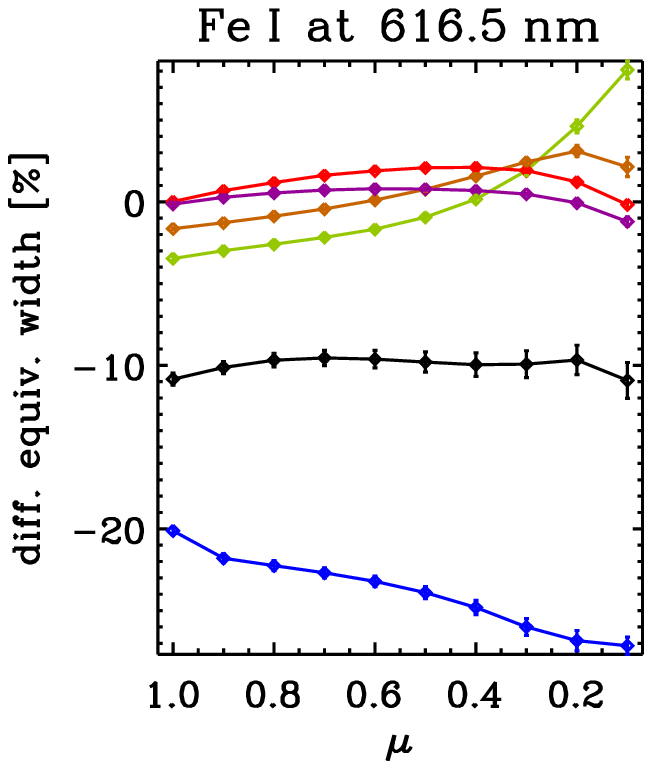}~~\includegraphics[width=4.1cm]{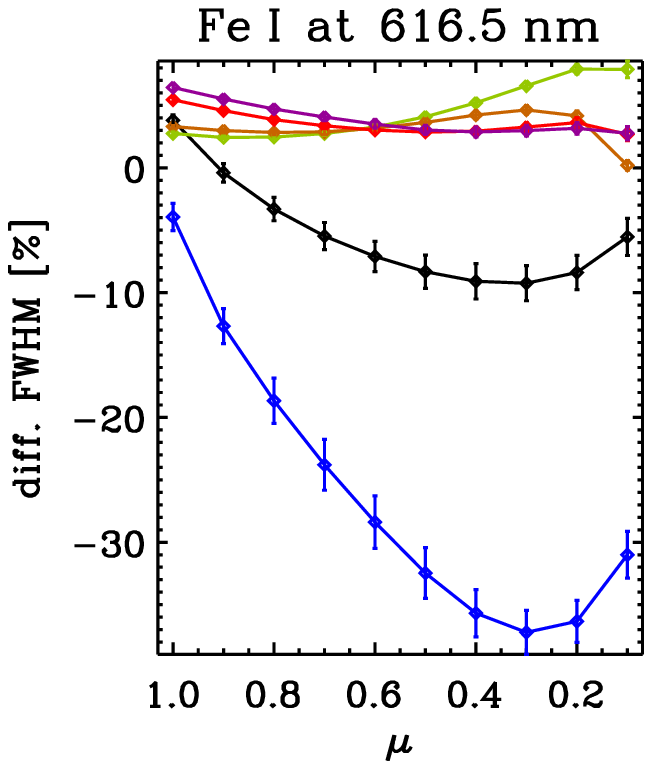}~~\includegraphics[width=4.1cm]{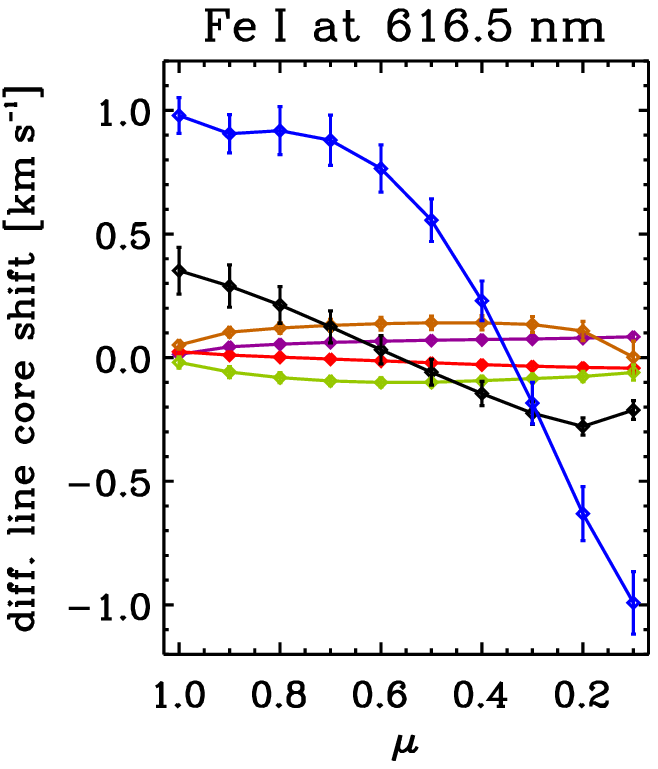}~~\includegraphics[width=4.1cm]{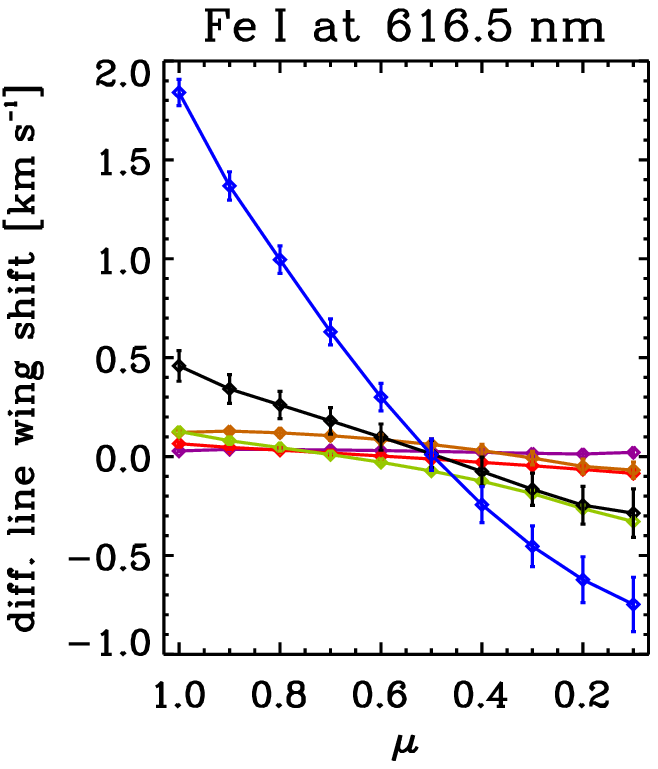}\\[2mm]
\includegraphics[width=4.1cm]{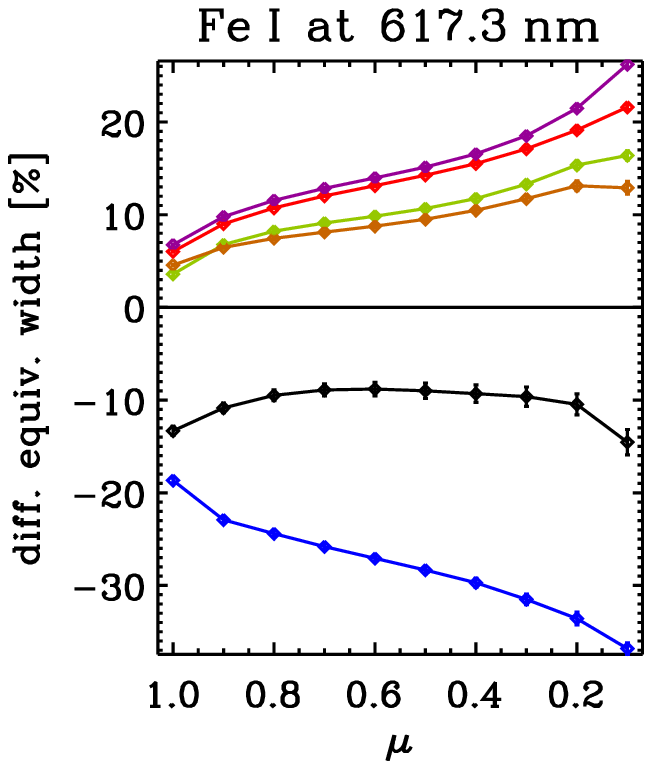}~~\includegraphics[width=4.1cm]{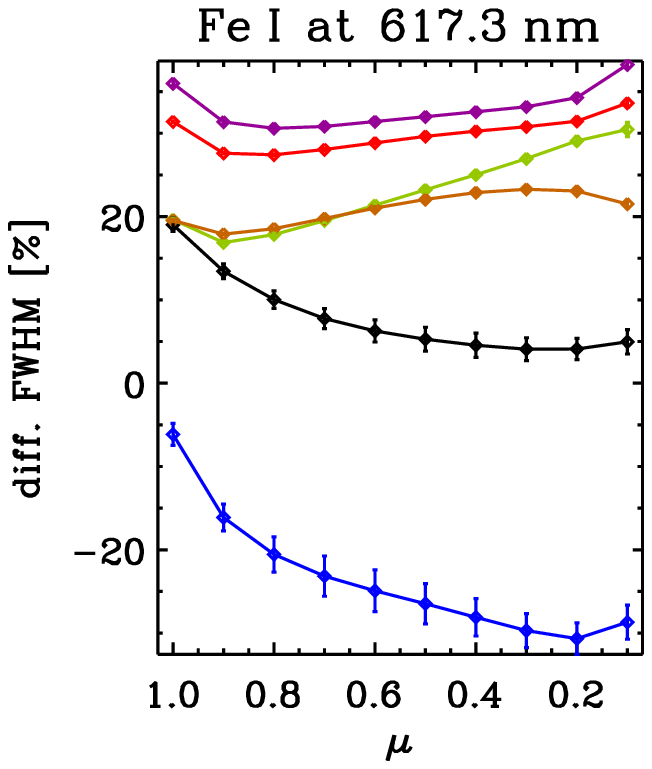}~~\includegraphics[width=4.1cm]{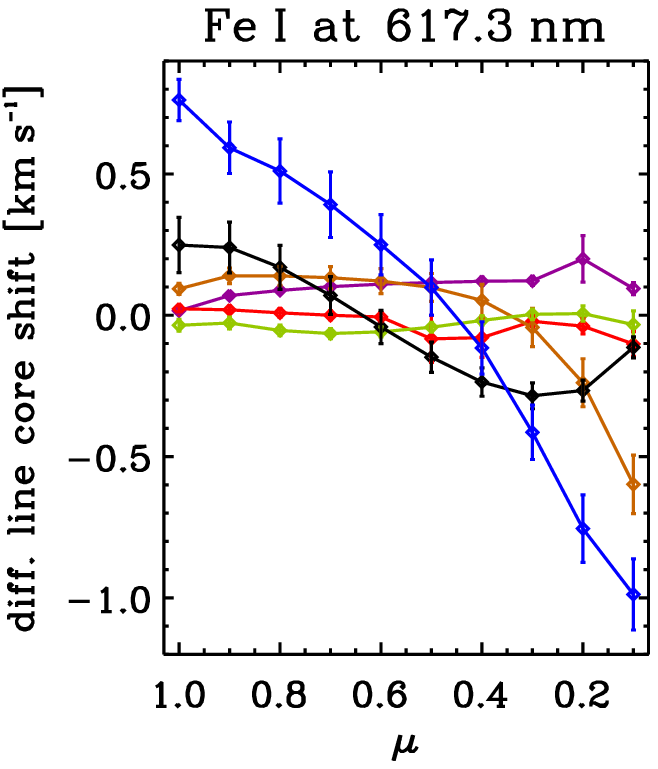}~~\includegraphics[width=4.1cm]{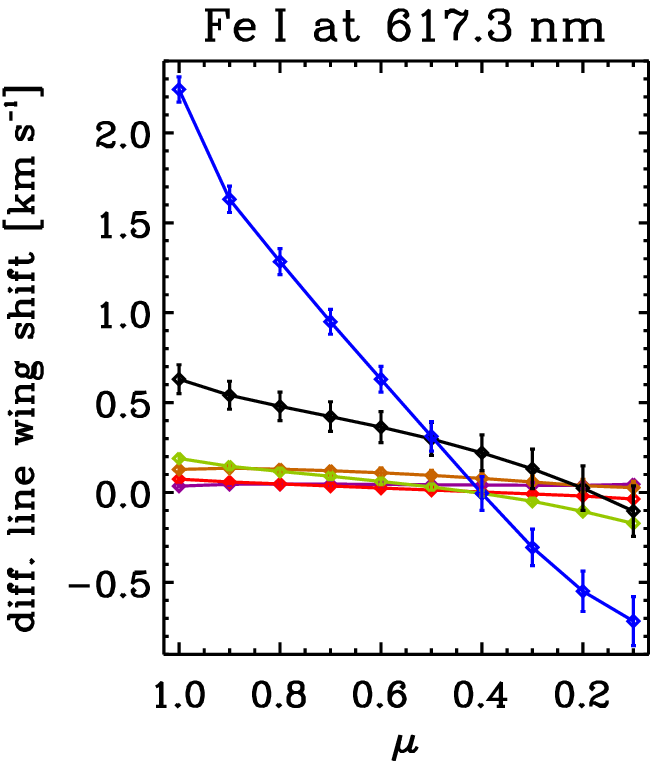} \\[2mm]
\includegraphics[width=4.1cm]{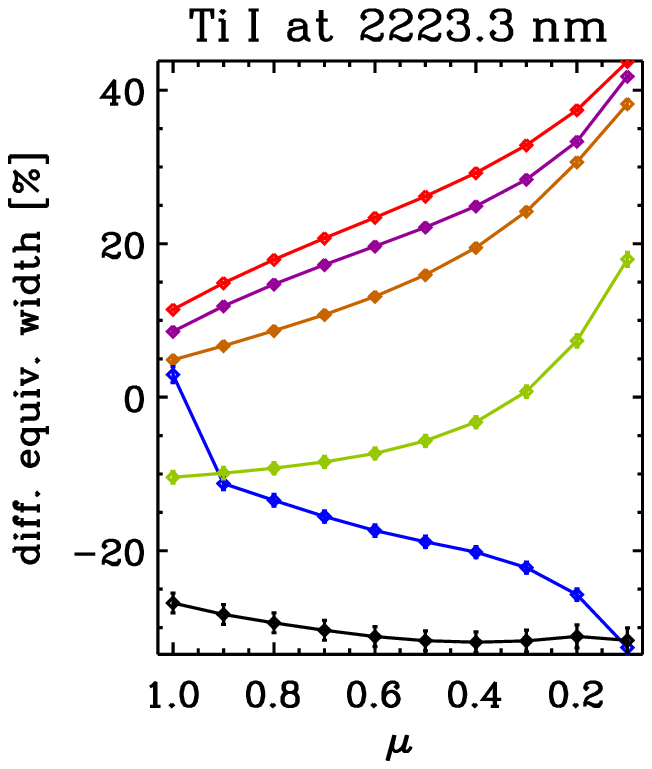}~~\includegraphics[width=4.1cm]{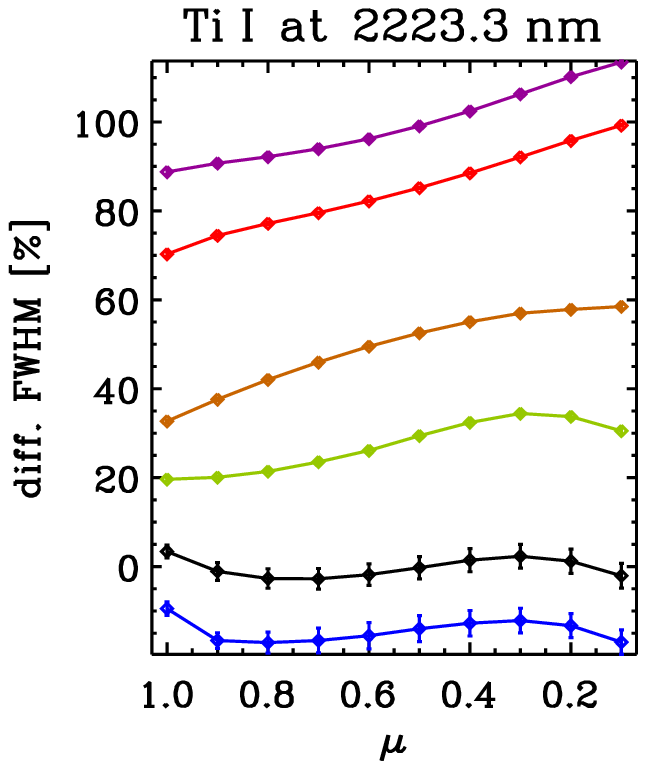}~~\includegraphics[width=4.1cm]{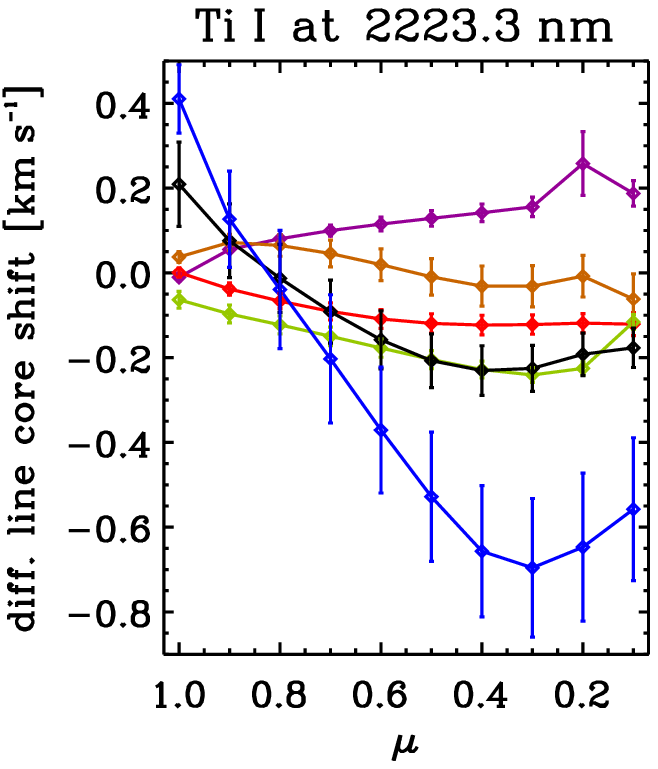}~~\includegraphics[width=4.1cm]{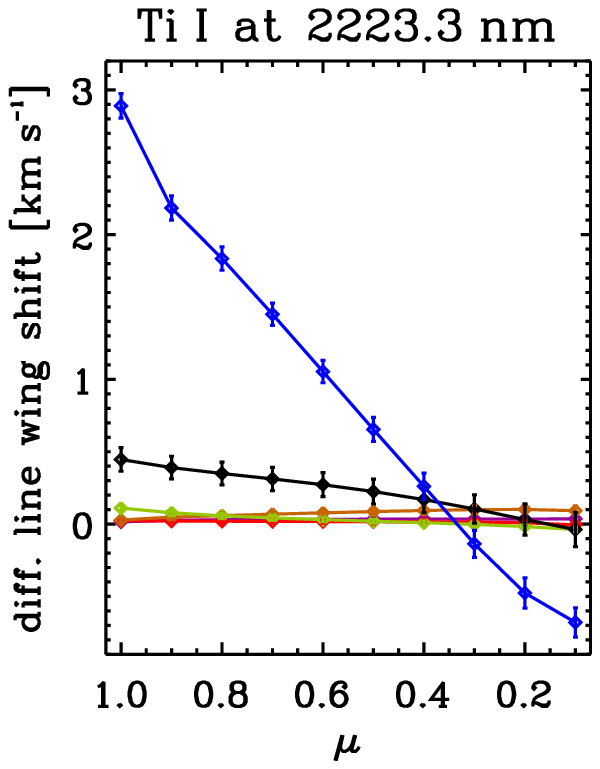}\\[1mm]
\includegraphics[width=5.8cm]{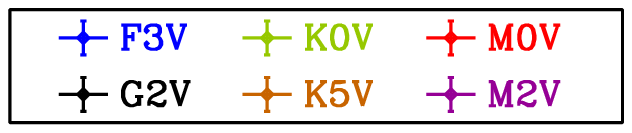}
\caption{Difference of equivalent width, FWHM, line core shift, and line wing shift ({\it from left to right}) between  simulations with $B_0=500\,\mathrm{G}$ and the respective non-magnetic simulations for the three spectral lines as a function of $\mu$.}\label{fig:lclv}
\end{figure*}
%
%
%
%
In Paper~II, the variation of the equivalent width (EW), the full width at
half maximum (FWHM), and the Doppler shifts of the cores and wings of spectral
lines (in particular of the Fe\,\textsc{i} line at 617.3\,nm) as a function of
viewing angle $\mu=\cos\vartheta$ were analysed for the non-magnetic
simulations. Here, we show the effect of the magnetic field on the
centre-to-limb variation of the line profiles. As discussed above for the
vertical (i.\,e. disc-centre) spectral line profiles, the magnetic field has a
substantial impact on the line profile shape
parameters. Figure~\ref{fig:lclv_F3} shows this for the Fe\,\textsc{i} line at
617.3\,nm for the example of the F3V star. The EW decreases with increasing
$B_0$ at all disc positions (i.\,e. for all values of $\mu$). This is mostly
caused by line weakening in the magnetic flux concentrations
(cf. Fig.~\ref{fig:local_spectra}). The centre-to-limb variation of the FWHM
does not differ much between the non-magnetic and the 20\,G and 100\,G
runs. At $B_0=500\,G$, however, the FWHM near the limb is considerably smaller
than in the non-magnetic run. This is a result of the stronger corrugation of
the optical surface in the magnetic run, which hides from view the part of the
granular flows receding from the observer as well as the cool downflows and
thus reduces the broadening induced by variations of the line-of-sight
velocity and temperature. The centre-to-limb variations of the effective
Doppler shifts of the line differ considerably between the runs. At the disc
centre, the line wings\footnote{Consistent with the definition in Paper~II,
  the line wing shift is determined as the displacement of the bisector from
  the rest wavelength of the spectral line at 5\% of the line depth.} are
strongly shifted to the red in the 500\,G runs owing to the higher downflow
velocities at constant optical depth (cf. Fig~\ref{fig:line_contribution}) in
this run in comparison to the non-magnetic runs. Near the limb, however, the
stronger corrugation in the 500\,G run results in an effective blue shift, as
the receding flows are mostly hidden behind the granules. As the effect of the
magnetic field on the shifts of line cores and wings is qualitatively and
quantitatively different, the magnetic field also has a strong impact on the line bisectors.\par
%
%
\begin{figure*}
\centering
\includegraphics[width=4.7cm]{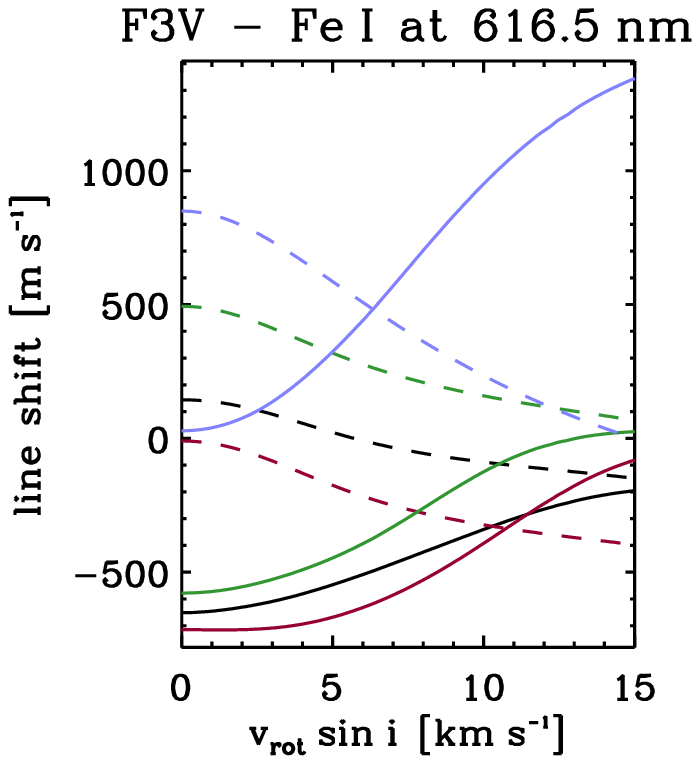}~\includegraphics[width=4.7cm]{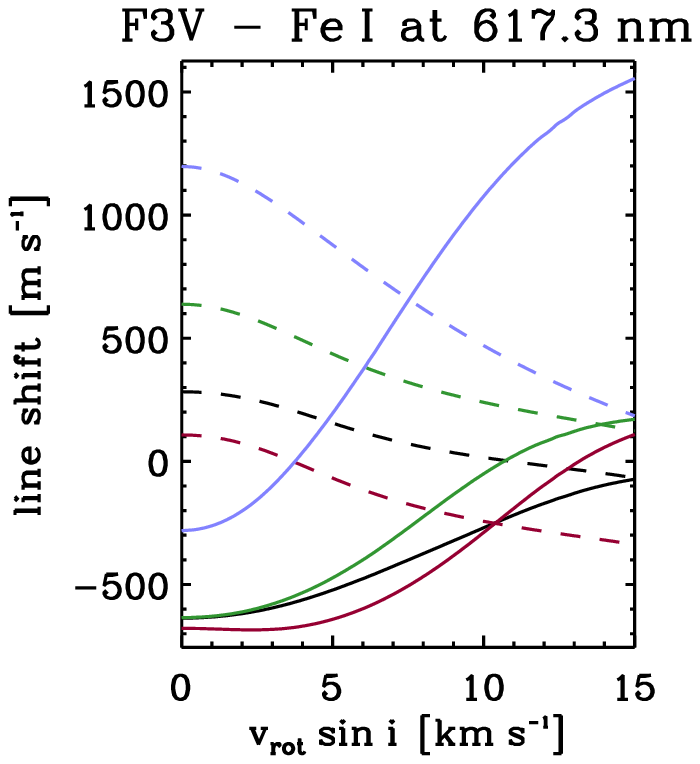}~\includegraphics[width=4.7cm]{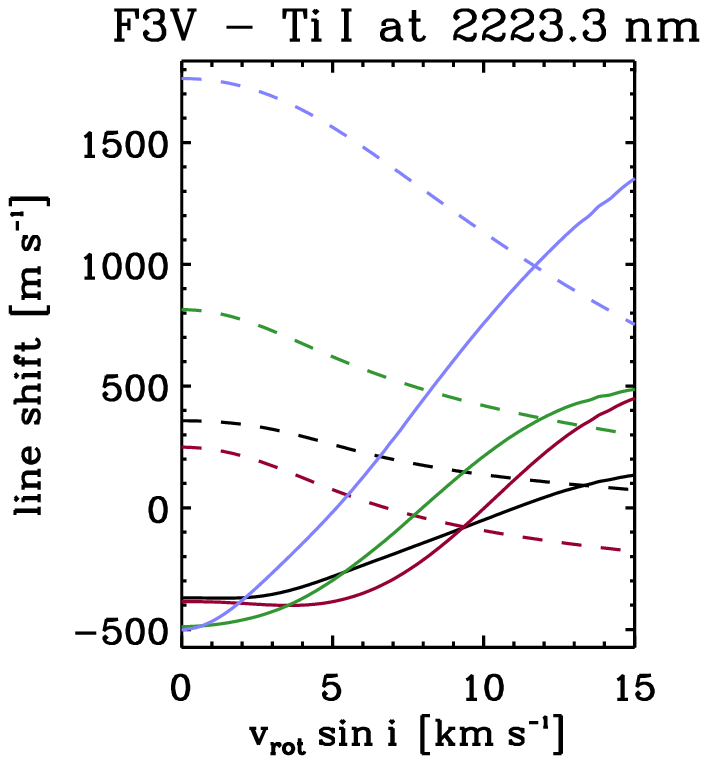}\\
\includegraphics[width=5.7cm]{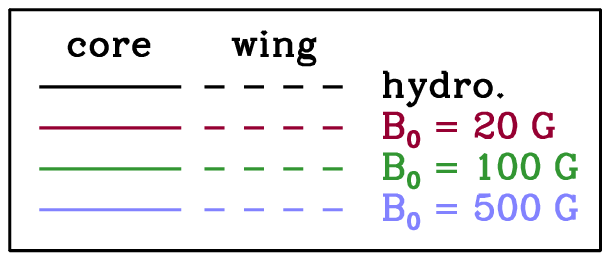}
\caption{Doppler shift of the line profile cores ({\it solid curves}) and wings ({\it dashed curves}) of the three spectral lines in the F3V simulations as function of $\vel_{\mathrm{rot}}\sin i$. In all cases, $i=60^{\circ}$ and $\alpha=0.2$ was assumed.}\label{fig:rotashift}
\end{figure*}
%
%
%
Figure~\ref{fig:lclv} shows the difference in EW, FWHM, line core shift, and
line wing shift between the 500\,G runs and the non-magnetic runs as a
function of $\mu$ for all spectral types and all three spectral lines (for the Fe\,\textsc{i} line at 617.3\,nm in the F3V star this corresponds to the difference between the blue and the black curve in Fig.~\ref{fig:lclv_F3}). The
differences in EW and FWHM are given as relative difference in per cent, while
the Doppler shifts are given as absolute differences in
$\mathrm{km\,s^{-1}}$. For both Fe\,\textsc{i} lines, the EW is smaller with
magnetic field than without in the G2V and F3V simulations
(cf. Fig.~\ref{fig:lclv_F3}) owing to line weakening in the bright magnetic
flux concentrations. In the other stars, line weakening does not play an
important role for these lines. Here, the EW is unaffected in the magnetically
insensitive Fe\,\textsc{i} line at 616.5\,nm while it is increased in the
magnetically sensitive line at 617.3\,nm owing to broadening by the Zeeman
effect, which reduces the degree of saturation of the line. For the
Ti\,\textsc{i} line, line weakening also affects the EW in the K0V star at
disc centre. We note that the EW of this line is very small in the F3V
simulation, particularly near the disc centre (see Paper~II) and only regions
with exceptional temperature gradients and high density contribute
substantially to the average profile, which might explain why the impact of
the magnetic field on the EW of this line is not similar or stronger than in
the G2V star as expected.\par
The FWHM follows a similar trend: for the hotter stars, the FWHM tends to be
smaller with magnetic field while, in the cooler stars, it is almost unchanged
in the magnetically insensitive line and strongly enhanced in the magnetically
sensitive lines. This again is showing the relative importance of line
weakening and Zeeman broadening. As the line weakening affects mainly the contributions from
downflows, it reduces the FWHM.\par 
The effective Doppler shifts in the lines show an impact of the magnetic field
which is almost independent of the sensitivity of the line to the magnetic
field. The disc centres of the F3V and G2V stars show strongly redshifted line
wings and slightly redshifted line cores in the 500\,G runs (relative to the
non-magnetic runs) in both Fe\,\textsc{i} lines. As shown earlier, this is
caused by a strongly enhanced red wing, resulting from the larger filling
factor of downflows (see Paper~III) and the higher downflow speed at the
optical depth range where these lines form. The Zeeman effect only slightly
enhances this effect by broadening the spectral line originating from the
magnetised downflows. Consequently, the relative redshifts in line wings and
cores are somewhat higher in the 617.3\,nm line than in the 616.5\,nm. Towards
the limb, these two stars show a small relative blueshift of line wings and
cores with respect to the non-magnetic runs (which is probably due to the stronger corrugation of the optical surface). In the cooler
stars, the impact of the magnetic field on the Doppler shifts of the line
cores and wings and on their dependence on $\mu$ is much weaker since the
convective velocities (particularly in the magnetic flux concentrations) are
considerably smaller.\par
%
%
%
%
\subsection{Disc-integrated spectral line profiles}\label{sec:discint}
Applying the numerical method outlined in Paper~II, disc-integrated spectral
line profiles were calculated using the angle dependent spectral line profiles
from the local simulation boxes. So far, we do not include any large-scale
surface inhomogeneities (such as active regions), but assume the stars to be
homogeneous on large scales (i.\,e. having the properties of the 3D simulation
with a constant $B_0$ across the entire disc). The disc integration includes
the rotation of the star, which follows the simple law given in
Eqn.~(\ref{eqn:difrot}) and thus introduces three parameters: the
equatorial velocity of the rotation, $\vel_{\mathrm{rot}}$, the inclination,
$i$, of the rotation axis with respect to the line of sight, and the
differential rotation parameter, $\alpha$.\par
%
%
\begin{figure*}
\centering
\includegraphics[width=7.1cm]{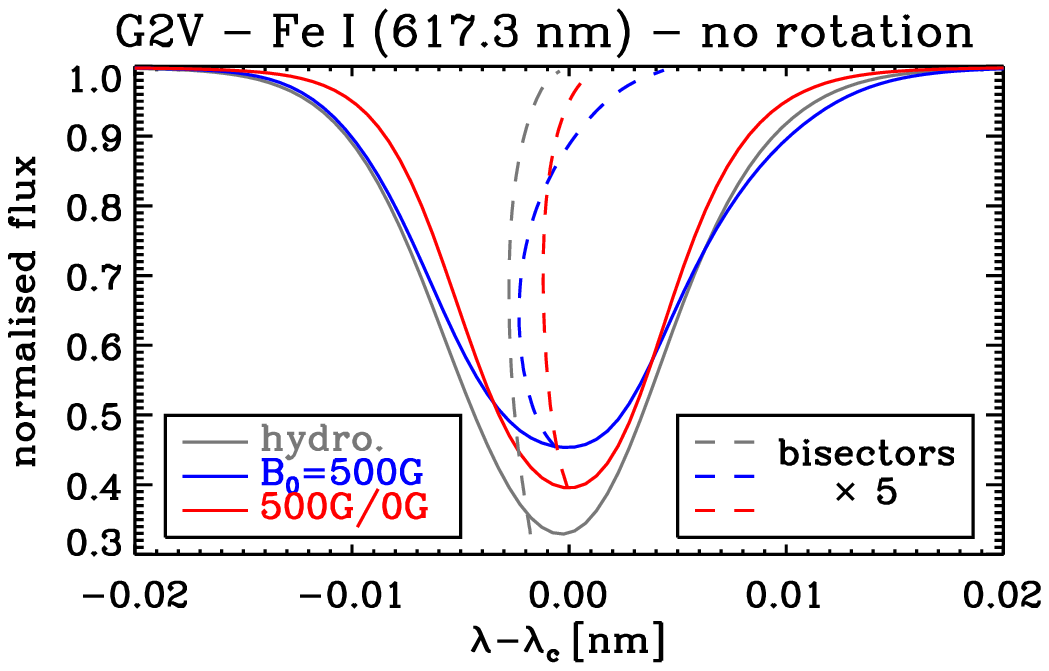}~\includegraphics[width=7.1cm]{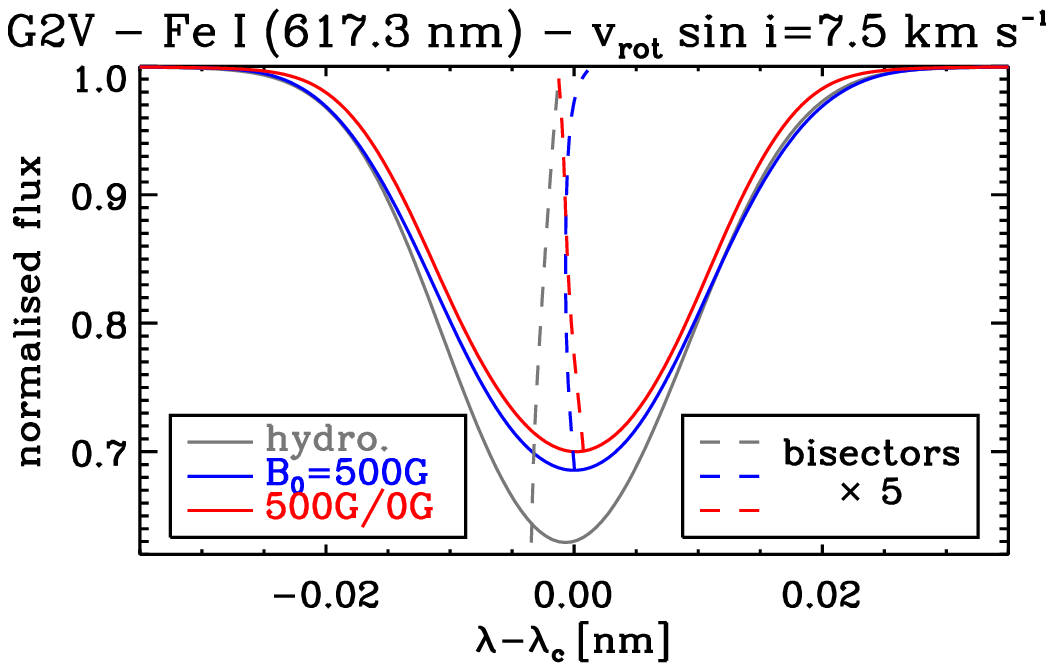}
\caption{Disc-integrated profiles of the Fe\,\textsc{i} line at 617.3\,nm without rotation ({\it left}) and with $\vel_{\mathrm{rot}}\,\sin i=7.5\,\mathrm{km\,s^{-1}}$ (differential, $\alpha=0.2$, $i=60^{\circ}$; {\it right}) for G2V simulations. The grey line represents the profile of the non-magnetic run ({\it hydro.}), the blue curve is the profile of the 500\,G run, the red curve is calculated using the thermodynamic structure of the 500\,G run but assuming $B\equiv 0\,\mathrm{G}$ for the line calculation (see text for further explanation). Dashed curves give the bisectors of the profiles with $\lambda-\lambda_{\mathrm{c}}$ values magnified by a factor of 5 for clarity. We note the different scales of ordinate and abscissa.}\label{fig:sample_TDvsZ}
\end{figure*}
%

%
As shown in Figs.~\ref{fig:lclv_F3} and~\ref{fig:lclv}, the 500\,G runs of the
hottest two stars of our sequence show redshifted line wings and (to a lesser
extent) redshifted line cores at the disc centre. In the disc-integrated
spectral line profiles, there is still a relative redshift with respect to the
non-magnetic simulations, but the shift is considerably smaller. These
effective Doppler shifts turn out to be sensitive to the rotational velocity
as the spectral line profiles get distorted by the rotational broadening
(cf. Paper~II). Figure~\ref{fig:rotashift} illustrates this for the F3V star
(where the shifts are largest) with the assumption of $i=60^{\circ}$ and
$\alpha=0.2$. In all three spectral lines, the line core moves redwards with
increasing rotational velocity and the line wings move bluewards. Although
this is the case in the non-magnetic and the magnetic runs alike, the effect
becomes stronger with magnetic field: in the 500\,G runs, the redshift in the
line core reaches values around $1.5\,\mathrm{km\,s^{-1}}$ at a
$\vel_{\mathrm{rot}}\sin i$ of $15\,\mathrm{km\,s^{-1}}$.\par
%
%
From the above (Sects.~\ref{sec:local} and \ref{sec:lclv}) it is evident that
the magnetic field has an impact on the spectral line profiles not only by
virtue of the Zeeman effect, but also owing to the local modifications of the
thermodynamical structure and the flow patterns caused by the magnetic flux
concentrations. In order to disentangle these two effects, the spectral line
synthesis for the 500\,G runs was carried out again ignoring the magnetic
field for the line synthesis. The result is indicated by the red curve in
Fig.~\ref{fig:sample_TDvsZ} for the example of the Fe\,\textsc{i} line at
617.3\,nm and the G2V-star simulation. For comparison, the line profile of the
non-magnetic simulation (grey) and of the full synthesis of the 500\,G run
(blue) are shown as well. Without rotation, the difference between the grey
and red curves, which is exclusively caused by the modifications in
thermodynamical structure and flows, is approximately equal in magnitude to
the change induced by the Zeeman effect (difference between blue and red
curves). While the local structure modifications shift the line somewhat to
the red and reduce its equivalent width and FWHM, as discussed above
(cf. Fig.\ref{fig:lclv}), the Zeeman effect broadens the line, as
expected. This broadening, however, is not exactly symmetric. The magnetic
field is predominantly located in downflows and is thus correlated with a
Doppler shift leading to a more broadened red wing. As a result the Zeeman effect increases the C-shape of the profile of this spectral line as can be seen in the bisectors (dashed curves in the figure). In the other stars and in the Ti\,\textsc{i} line a qualitatively similar impact on the bisectors is visible (see Figs.~\ref{fig:sample_app1} and~ \ref{fig:sample_app2}), albeit less pronounced.\par
At a moderate rotational velocity of $7.5\,\mathrm{km\,s^{-1}}$ (see right panel of Fig.~\ref{fig:sample_TDvsZ}), the
broadening due to the Zeeman effect is already very small compared to the
rotational broadening. Consequently, the effect of the modified flows and thermodynamical
structure on the line is overall much larger than the Zeeman effect.\par
This calculation has been carried out for all simulations and all spectral lines (more plots are given in Appendix~\ref{app:figs}). We find the locally
modified thermodynamics and flows to have a major impact on the line profiles
in the F-, G-, and K-type stars. Only in the M-star simulation is the effect of
thermodynamics significantly smaller than that of the Zeeman effect. This
has several reasons. Firstly, line weakening does not play a role here for the
spectral lines considered. Secondly, the absolute contrast of the temperature
is smaller than in the hotter stars. Thirdly, the convective velocities are
generally smaller; in particular, there are no large downflow velocities in
the magnetic structures as they are found in the hotter stars.

\section{Concluding remarks}
We have shown that the intensity, its contrast and centre-to-limb variation as
well as spectral line profile shapes are strongly influenced be a moderate
local magnetic field. In unipolar regions with an average magnetic field of up
to 500\,G, the magnetic flux is strongly concentrated in localised structures
of reduced or enhanced intensity. F-, G-, and K-type stars show more magnetic
flux concentrations of enhanced intensity under these conditions, whereas
M-type stars have prominent dark pore-like structures. These structures
generally become bright near the limb and thus reduce the limb darkening
(magnetic limb brightening). Moreover, the intensity contrast is enhanced,
particularly near the limb. The influence on spectral line profiles is not
only caused by the action of the Zeeman effect on the magnetically sensitive
lines, but also by the presence of the magnetic flux concentrations which
substantially deviate in their thermodynamical structure and flow patterns
from the non-magnetised surroundings. Moreover, the height dependence of the
magnetic field and the flow velocity as well as convective collapse events
(resulting in strong downflows in magnetic flux concentrations) add
asymmetries to the full-disc spectral line profiles, which can strongly depend
on the rotation velocity of star.\par For the synthesis of spectral lines as
well as for the calculation of the limb darkening, we have implicitly made the
assumption that the stars are completely covered with a unipolar field, which
is homogeneous on large scales.\footnote{As we only consider the Stokes-$I$
  component, it is not an important issue that such a global configuration is
  not divergence-free.} As the solar example as well as ZDI results show,
stellar magnetic fields are not only structured on small scales, but also
possess global structure (e.\,g. active latitudes, star spot groups, polar
fields). Even though a homogeneously magnetised star as we implicitly
assume does probably not exist in reality, the results presented in this paper
illustrate that the impact even of a moderate magnetic field on observable
quantities is expected to be significant. For Stokes-$I$ measurements the
Zeeman-effect signature can be largely obliterated by the effect of the
deviating atmospheric structure in the magnetised regions. Consequently,
magnetic field measurements for stars (at least for solar-like stars) have to
be taken with caution because the magnetic field is usually considered to
have no impact on the spectral lines apart from the Zeeman effect. The
comparison of magnetically sensitive and insensitive lines \citep[see
  e.\,g.][]{Robinson80} only helps to a certain extent, as this does not
account for the correlation of the magnetic field with, e.\,g. the velocity
field, which can modify the impact of the Zeeman effect.\par
In this paper, we do not consider spectral line profiles in polarised
light. The line weakening discussed here does also affect the other
Stokes components of the spectral lines. Moreover, a strong line-of-sight
velocity variation in the magnetised surface components (as we have found it at least for the
F3V star) will also affect the Stokes-V signal and reduce the resolution of
ZDI (which makes use of the line-of-sight velocity to obtain information on
the spatial distribution of the magnetic flux).\par
Great progress has been made in the detection and measurement of stellar
magnetic fields in recent years. To improve the methods it will be important to include the various effects of the magnetic field on
the stellar atmospheres resulting from its 3D structure.
\begin{acknowledgements}
  The authors acknowledge research funding by the Deutsche
  Forschungsgemeinschaft (DFG) under the grant SFB 963/1, project A16. AR has
  received research funding from the DFG under DFG 1664/9-2.
\end{acknowledgements}
\bibliography{paper4}

\begin{thebibliography}{26}
\expandafter\ifx\csname natexlab\endcsname\relax\def\natexlab#1{#1}\fi

\bibitem[{{Anders} \& {Grevesse}(1989)}]{AnGr89}
{Anders}, E. \& {Grevesse}, N. 1989, \gca, 53, 197

\bibitem[{{Baliunas} {et~al.}(1995){Baliunas}, {Donahue}, {Soon}, {Horne},
  {Frazer}, {Woodard-Eklund}, {Bradford}, {Rao}, {Wilson}, {Zhang}, {Bennett},
  {Briggs}, {Carroll}, {Duncan}, {Figueroa}, {Lanning}, {Misch}, {Mueller},
  {Noyes}, {Poppe}, {Porter}, {Robinson}, {Russell}, {Shelton}, {Soyumer},
  {Vaughan}, \& {Whitney}}]{MtWilson}
{Baliunas}, S.~L., {Donahue}, R.~A., {Soon}, W.~H., {et~al.} 1995, \apj, 438,
  269

\bibitem[{{Basri} {et~al.}(1990){Basri}, {Valenti}, \& {Marcy}}]{Basri1990}
{Basri}, G., {Valenti}, J.~A., \& {Marcy}, G.~W. 1990, \apj, 360, 650

\bibitem[{{Beeck} {et~al.}(2013{\natexlab{a}}){Beeck}, {Cameron}, {Reiners}, \&
  {Sch{\"u}ssler}}]{paper1}
{Beeck}, B., {Cameron}, R.~H., {Reiners}, A., \& {Sch{\"u}ssler}, M.
  2013{\natexlab{a}}, \aap, 558, A48

\bibitem[{{Beeck} {et~al.}(2013{\natexlab{b}}){Beeck}, {Cameron}, {Reiners}, \&
  {Sch{\"u}ssler}}]{paper2}
{Beeck}, B., {Cameron}, R.~H., {Reiners}, A., \& {Sch{\"u}ssler}, M.
  2013{\natexlab{b}}, \aap, 558, A49

\bibitem[{{Beeck} {et~al.}(2015){Beeck}, {Sch{\"u}ssler}, {Cameron}, \&
  {Reiners}}]{paper3}
{Beeck}, B., {Sch{\"u}ssler}, M., {Cameron}, R.~H., \& {Reiners}, A. 2015,
  submitted to \aap

\bibitem[{{Bessell}(1990)}]{Bessell90}
{Bessell}, M.~S. 1990, \pasp, 102, 1181

\bibitem[{Charbonneau(2010)}]{Charbonneau10}
Charbonneau, P. 2010, Liv. Rev. Sol. Phys., 7, 3

\bibitem[{{Donati}(2011)}]{Donati11}
{Donati}, J.-F. 2011, in IAU Symposium, Vol. 271, IAU Symposium, ed. N.~H.
  {Brummell}, A.~S. {Brun}, M.~S. {Miesch}, \& Y.~{Ponty}, 23

\bibitem[{{Frutiger}(2000)}]{spinor}
{Frutiger}, C. 2000, Diss. ETH, 13896

\bibitem[{{Frutiger} {et~al.}(2000){Frutiger}, {Solanki}, {Fligge}, \&
  {Bruls}}]{spinor2}
{Frutiger}, C., {Solanki}, S.~K., {Fligge}, M., \& {Bruls}, J.~H.~M.~J. 2000,
  \aap, 358, 1109

\bibitem[{{Jeffers} {et~al.}(2013){Jeffers}, {Barnes}, {Jones}, \&
  {Pinfield}}]{Jeffersetal2013}
{Jeffers}, S.~V., {Barnes}, J.~R., {Jones}, H., \& {Pinfield}, D. 2013, in
  European Physical Journal Web of Conferences, Vol.~47, European Physical
  Journal Web of Conferences, 9002

\bibitem[{{Johnson} \& {Morgan}(1951)}]{Johnson}
{Johnson}, H.~L. \& {Morgan}, W.~W. 1951, \apj, 114, 522

\bibitem[{{Keller} {et~al.}(2004){Keller}, {Sch{\"u}ssler}, {V{\"o}gler}, \&
  {Zakharov}}]{Keller04}
{Keller}, C.~U., {Sch{\"u}ssler}, M., {V{\"o}gler}, A., \& {Zakharov}, V. 2004,
  \apj, 607, L59

\bibitem[{{Kurucz}(1993)}]{atlas9}
{Kurucz}, R. 1993, ATLAS9 Stellar Atmosphere Programs and 2 km/s grid.~Kurucz
  CD-ROM No.~13.~ Cambridge, Mass.: Smithsonian Astrophysical Observatory,
  1993., 13

\bibitem[{{Morin} {et~al.}(2010){Morin}, {Donati}, {Petit}, {Delfosse},
  {Forveille}, \& {Jardine}}]{Morin10}
{Morin}, J., {Donati}, J.-F., {Petit}, P., {et~al.} 2010, \mnras, 407, 2269

\bibitem[{{Reiners}(2012)}]{Ansgar12}
{Reiners}, A. 2012, Liv. Rev. Sol. Phys., 9, 1

\bibitem[{{Reiners} {et~al.}(2014){Reiners}, {Sch{\"u}ssler}, \&
  {Passegger}}]{Ansgar2014}
{Reiners}, A., {Sch{\"u}ssler}, M., \& {Passegger}, V.~M. 2014, \apj, 794, 144

\bibitem[{{Robinson}(1980)}]{Robinson80}
{Robinson}, Jr., R.~D. 1980, \apj, 239, 961

\bibitem[{{Rogers}(1994)}]{opal2}
{Rogers}, F.~J. 1994, in IAU Colloq. 147: The Equation of State in
  Astrophysics, ed. G.~{Chabrier} \& E.~{Schatzman}, 16

\bibitem[{{Rogers} {et~al.}(1996){Rogers}, {Swenson}, \& {Iglesias}}]{opal1}
{Rogers}, F.~J., {Swenson}, F.~J., \& {Iglesias}, C.~A. 1996, \apj, 456, 902

\bibitem[{{Ros{\'e}n} \& {Kochukhov}(2012)}]{Rosenetal12}
{Ros{\'e}n}, L. \& {Kochukhov}, O. 2012, \aap, 548, A8

\bibitem[{{Saar} \& {Solanki}(1992)}]{SaSo92}
{Saar}, S.~H. \& {Solanki}, S.~K.~. 1992, in Astronomical Society of the
  Pacific Conference Series, Vol.~26, Cool Stars, Stellar Systems, and the Sun,
  ed. M.~S. {Giampapa} \& J.~A. {Bookbinder}, 259

\bibitem[{{Shulyak} {et~al.}(2014){Shulyak}, {Reiners}, {Seemann}, {Kochukhov},
  \& {Piskunov}}]{Denis14}
{Shulyak}, D., {Reiners}, A., {Seemann}, U., {Kochukhov}, O., \& {Piskunov}, N.
  2014, \aap, 563, A35

\bibitem[{{V{\"o}gler}(2003)}]{MURaM1}
{V{\"o}gler}, A. 2003, PhD thesis, Gerog-August-Universit{\"a}t G{\"o}ttingen

\bibitem[{{V{\"o}gler} {et~al.}(2005){V{\"o}gler}, {Shelyag}, {Sch{\"u}ssler},
  {Cattaneo}, {Emonet}, \& {Linde}}]{MURaM2}
{V{\"o}gler}, A., {Shelyag}, S., {Sch{\"u}ssler}, M., {et~al.} 2005, \aap, 429,
  335

\end{thebibliography}
\appendix 
\section{Additional figures}\label{app:figs}
\begin{figure*}
\centering
\includegraphics[width=7.1cm]{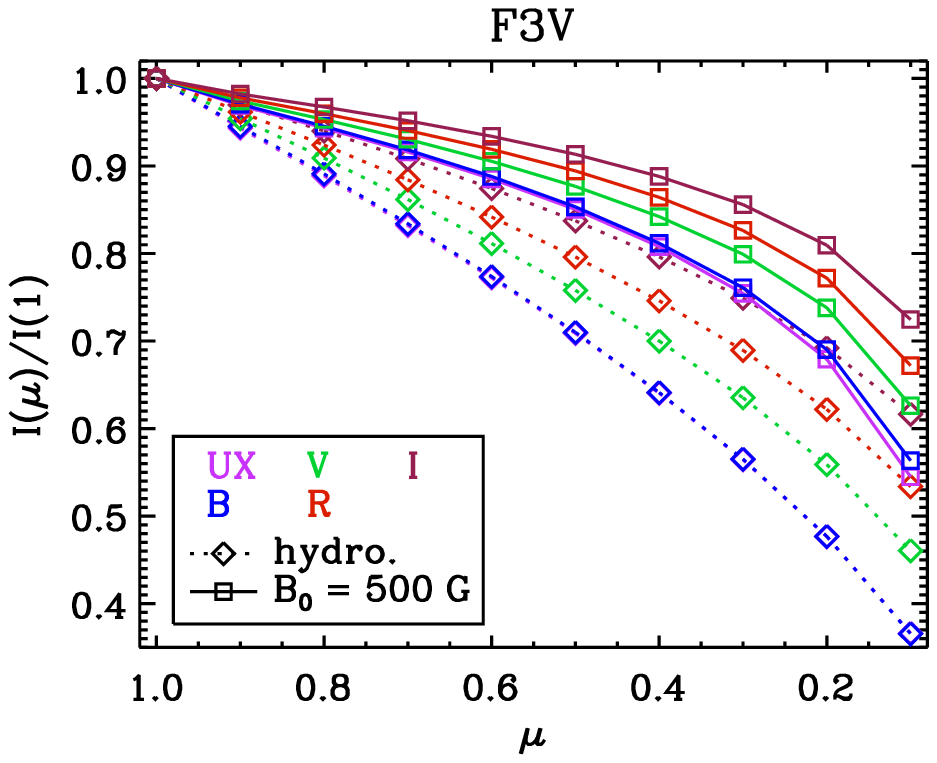}~\includegraphics[width=7.1cm]{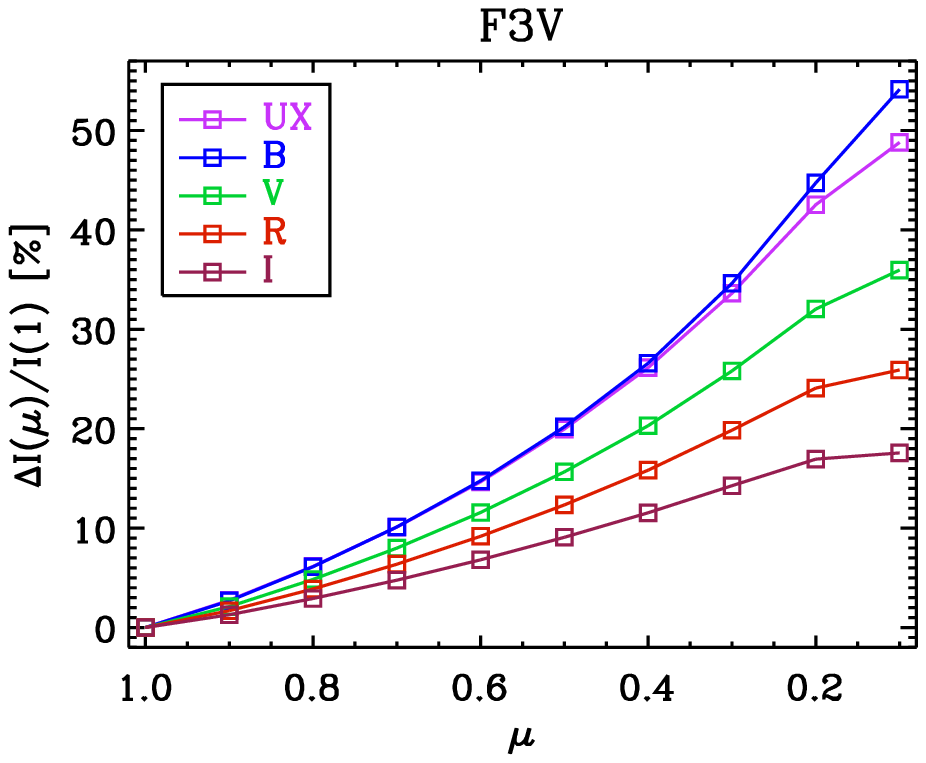}\\
\includegraphics[width=7.1cm]{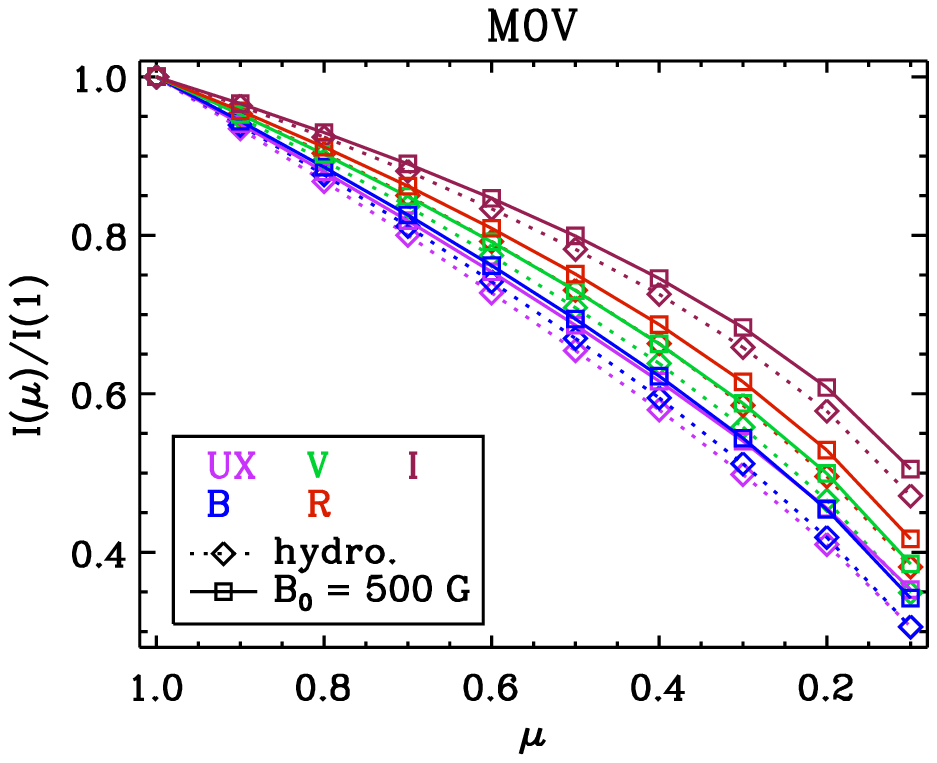}~\includegraphics[width=7.1cm]{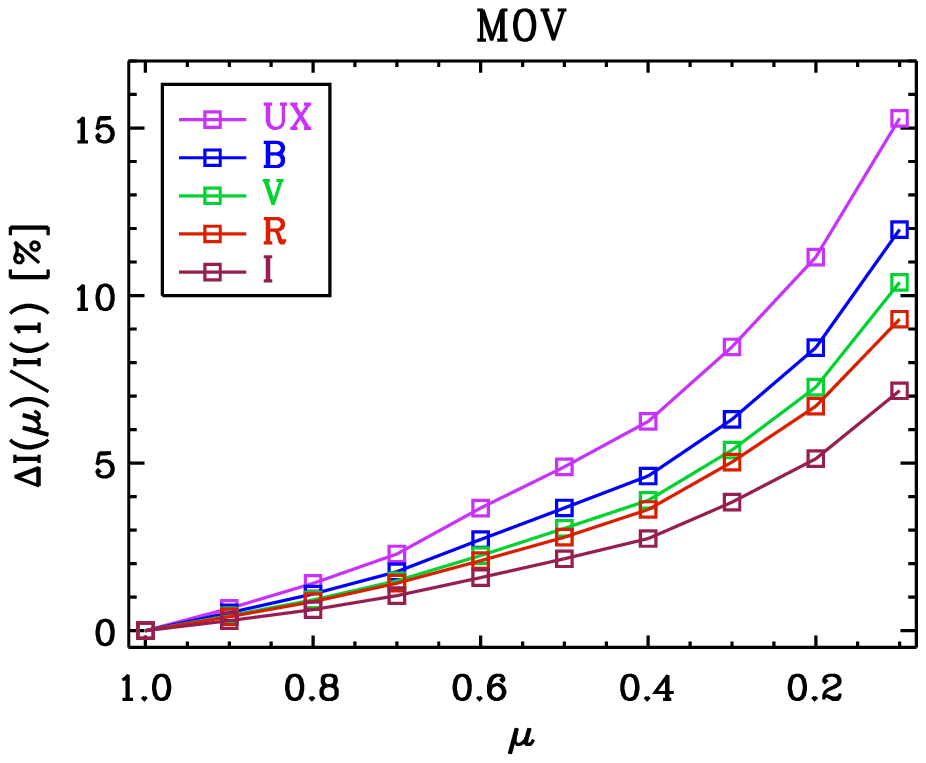}\\
\caption{Same as Fig.~\ref{fig:LD_Johnson} but for the F3V and the M0V star.}\label{fig:LD_app}
\end{figure*}

\begin{figure*}
\centering
\includegraphics[width=7.1cm]{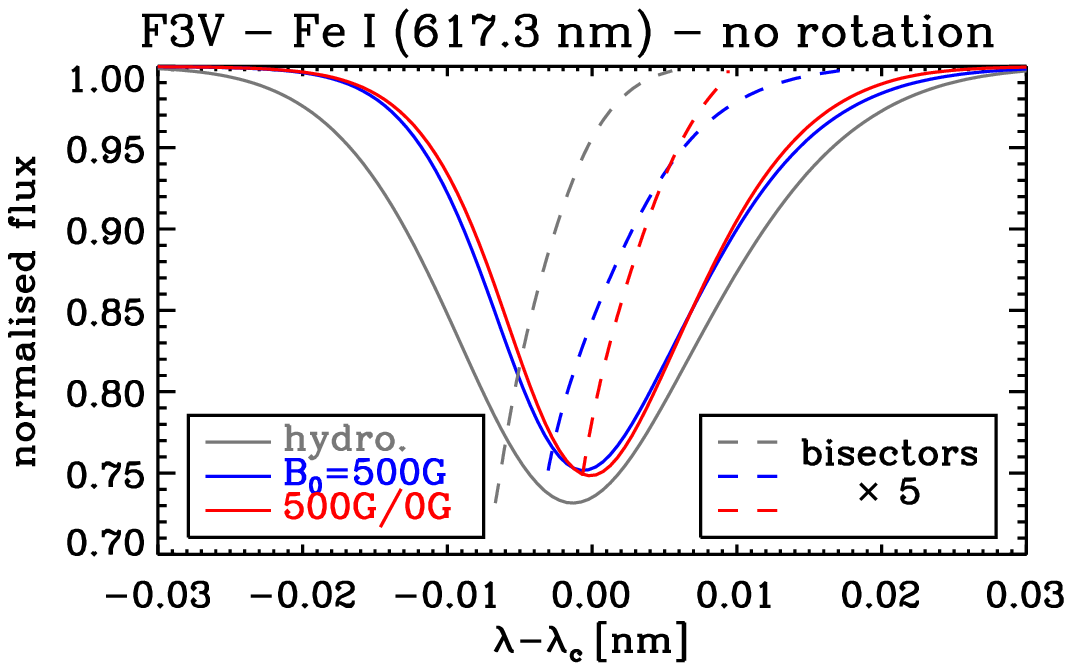}~\includegraphics[width=7.1cm]{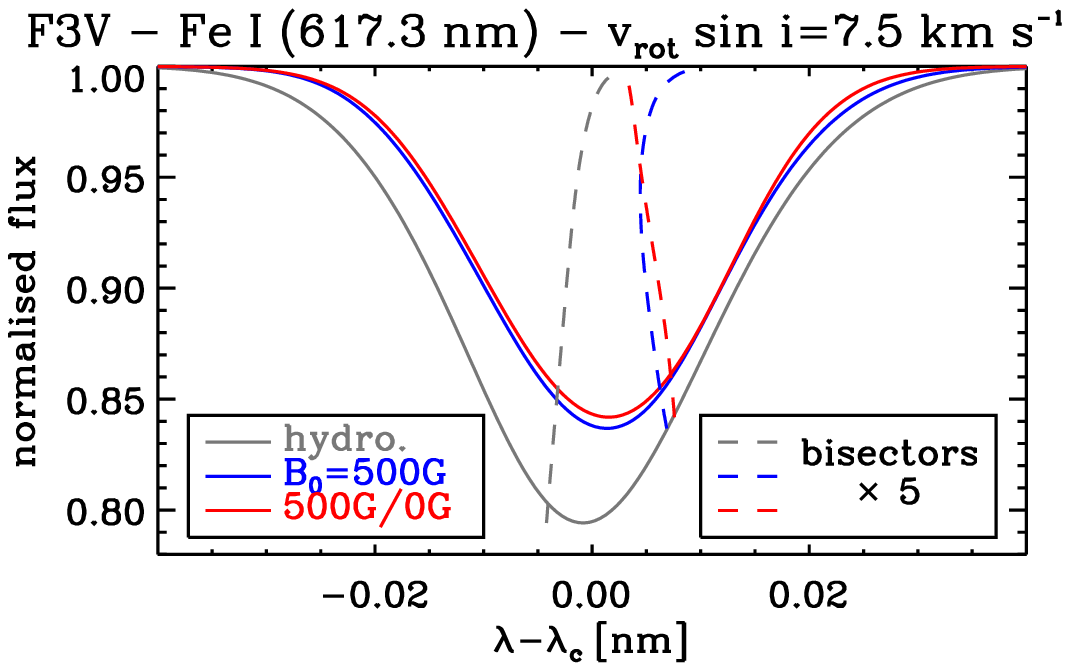}
\includegraphics[width=7.1cm]{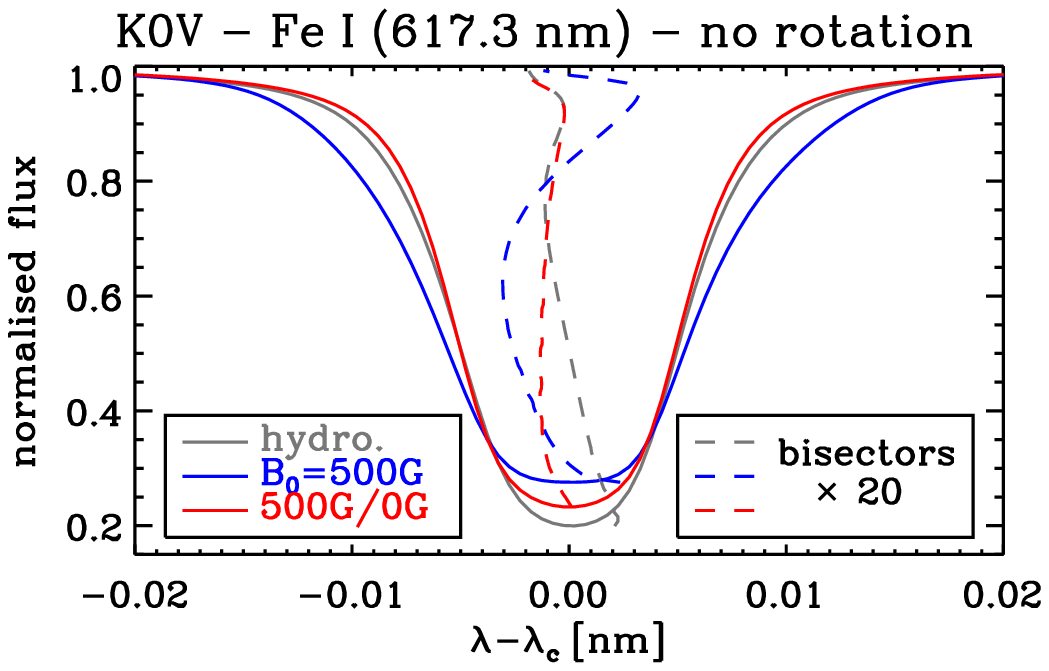}~\includegraphics[width=7.1cm]{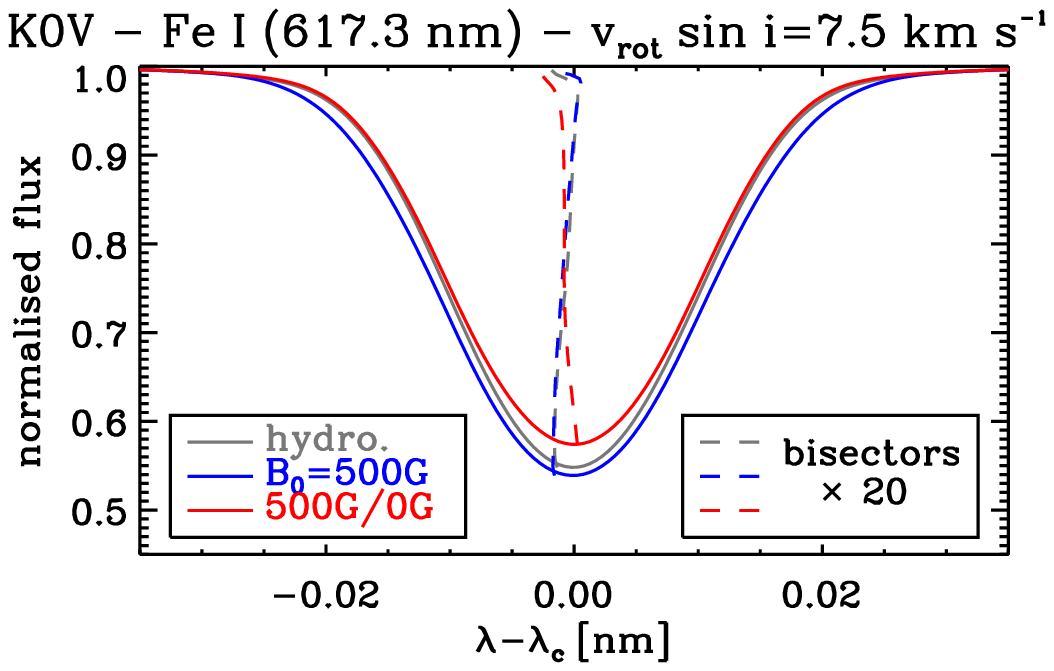}
\includegraphics[width=7.1cm]{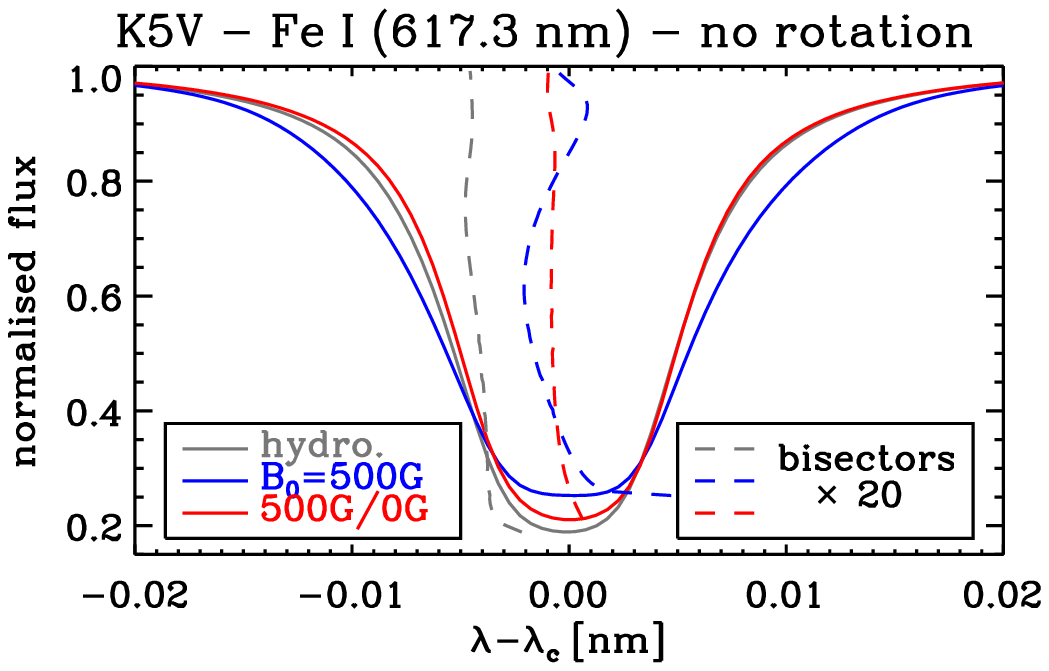}~\includegraphics[width=7.1cm]{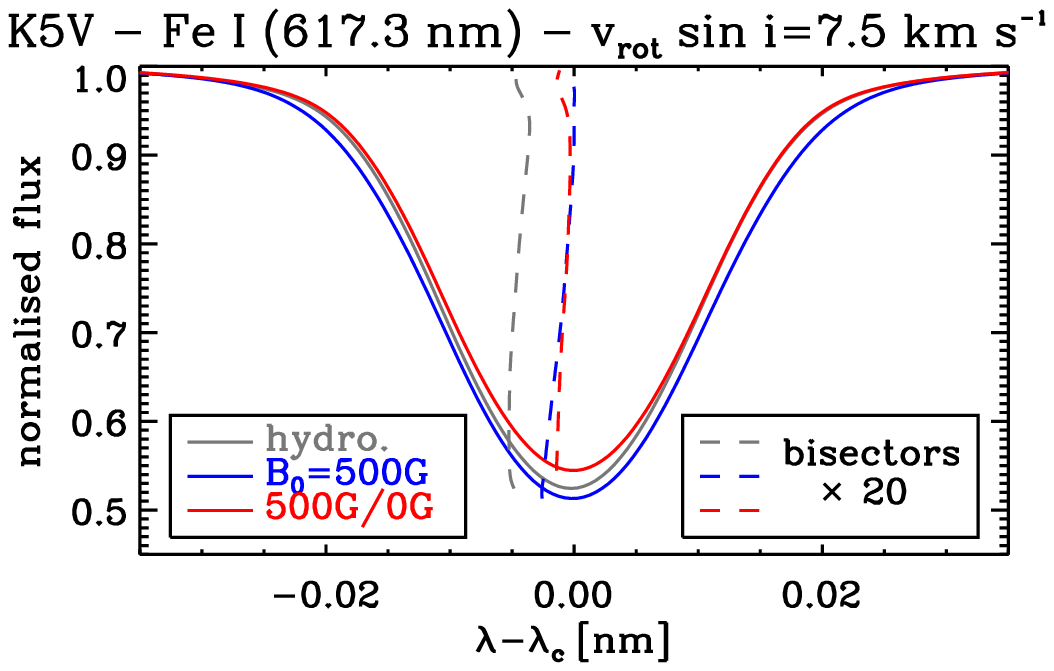}
\includegraphics[width=7.1cm]{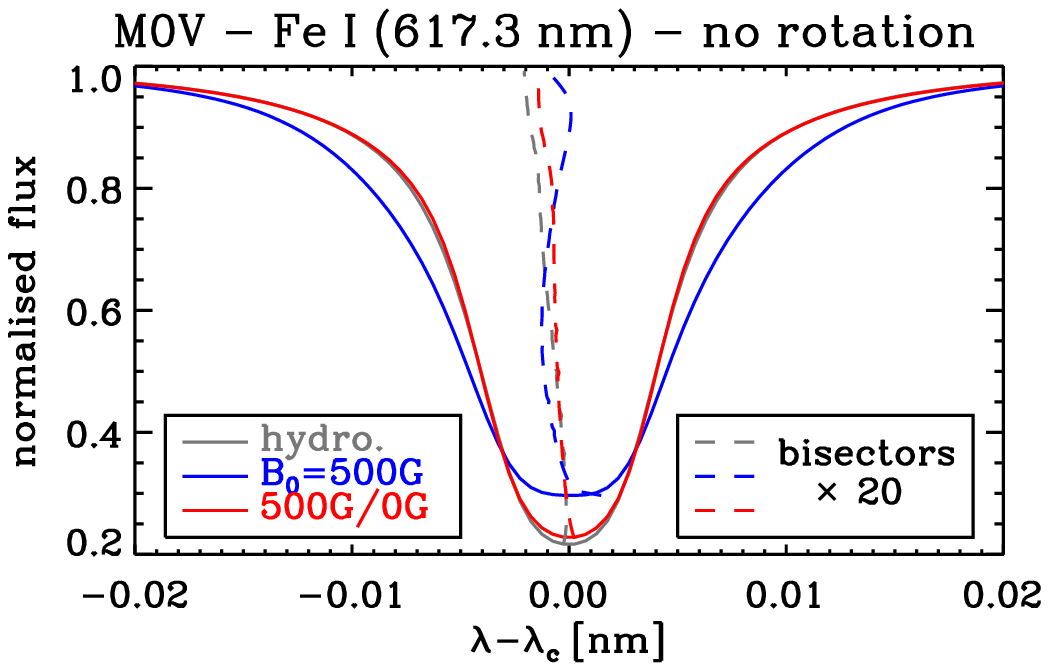}~\includegraphics[width=7.1cm]{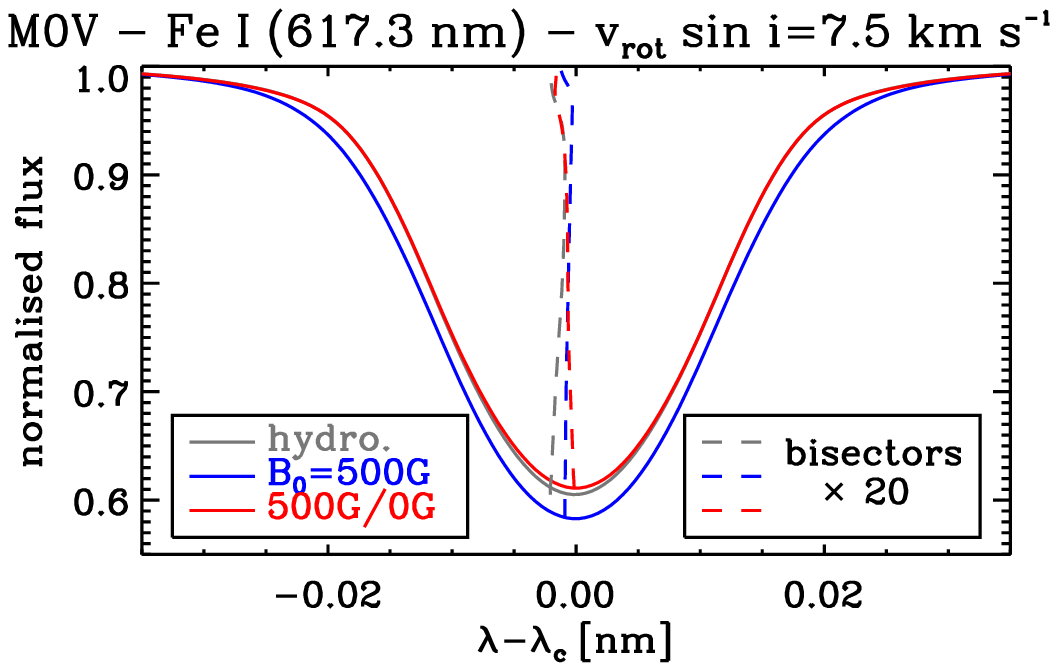}
\caption{Same as Fig.~\ref{fig:sample_TDvsZ}, but for the F3V, K0V, K5V and M0V simulations.}\label{fig:sample_app1}
\end{figure*}
\begin{figure*}
\centering
\includegraphics[width=7.1cm]{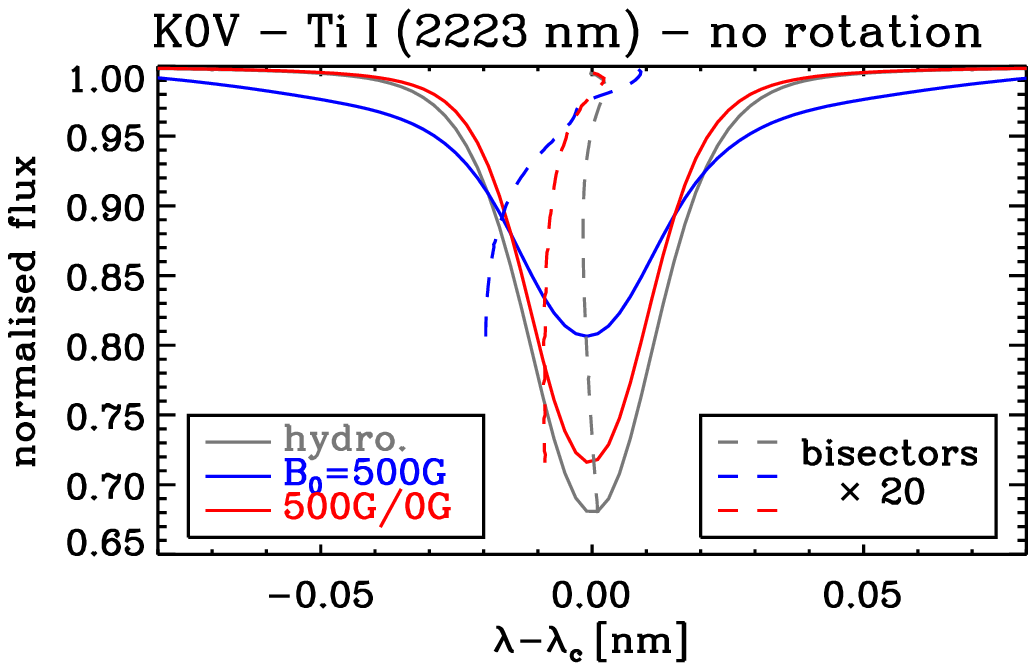}~\includegraphics[width=7.1cm]{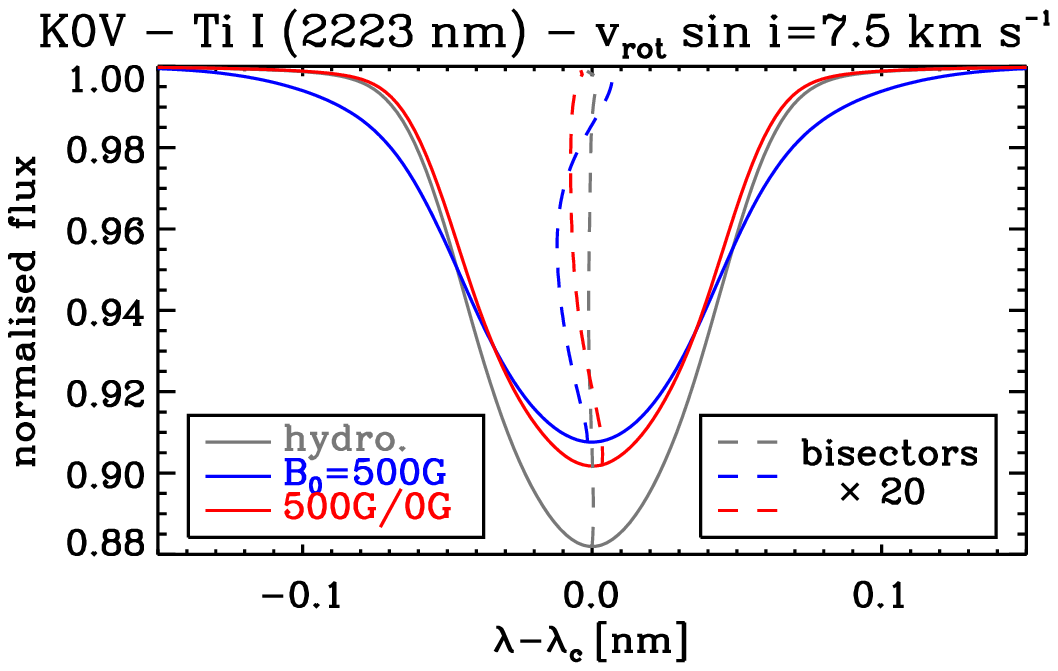}
\includegraphics[width=7.1cm]{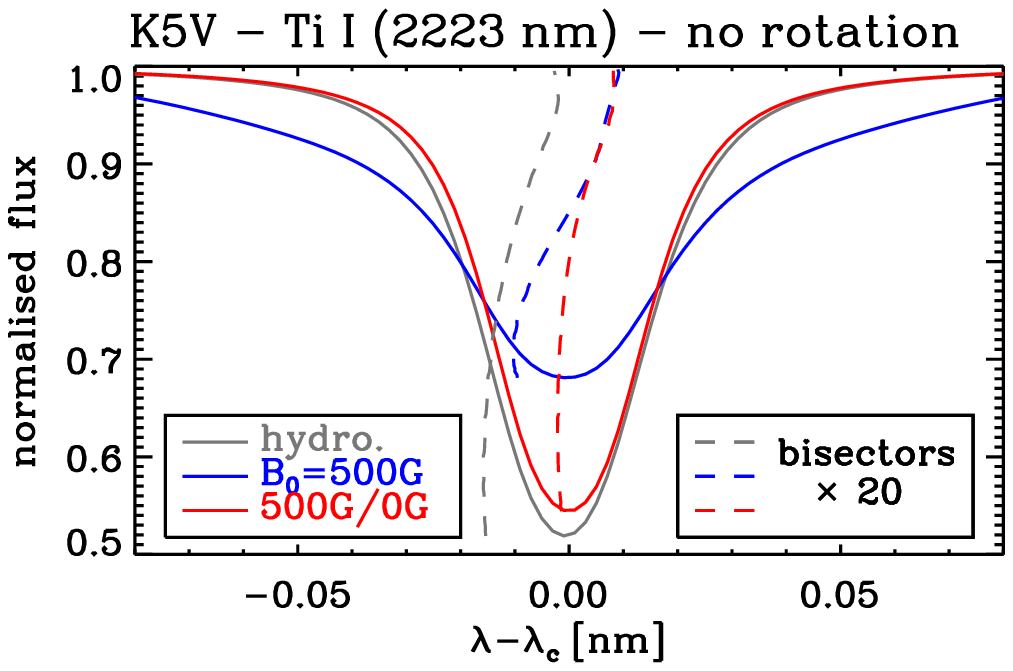}~\includegraphics[width=7.1cm]{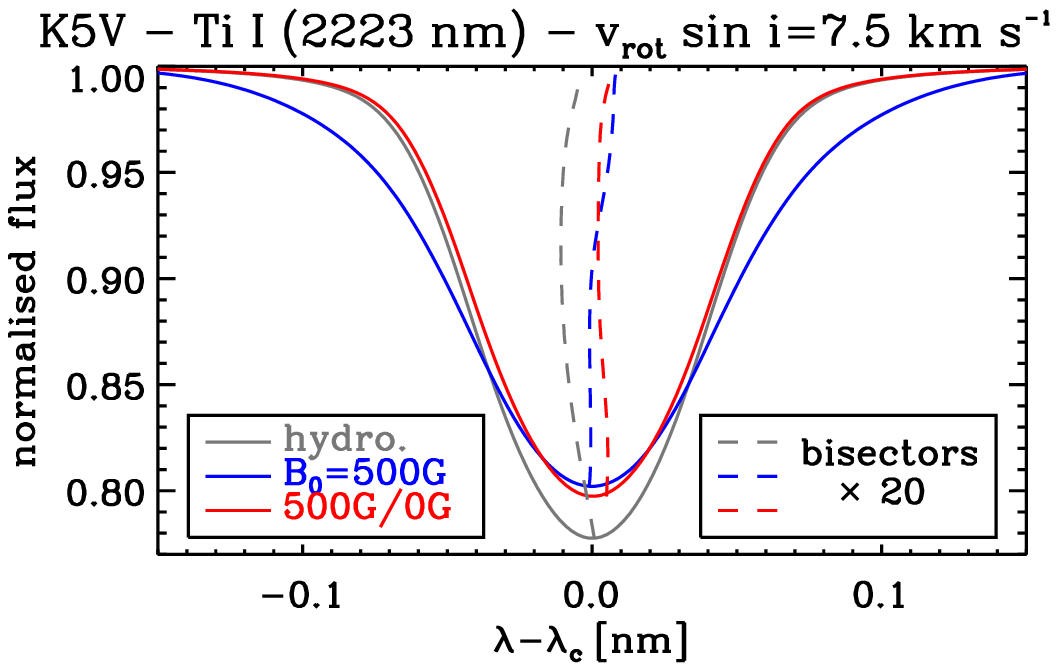}
\includegraphics[width=7.1cm]{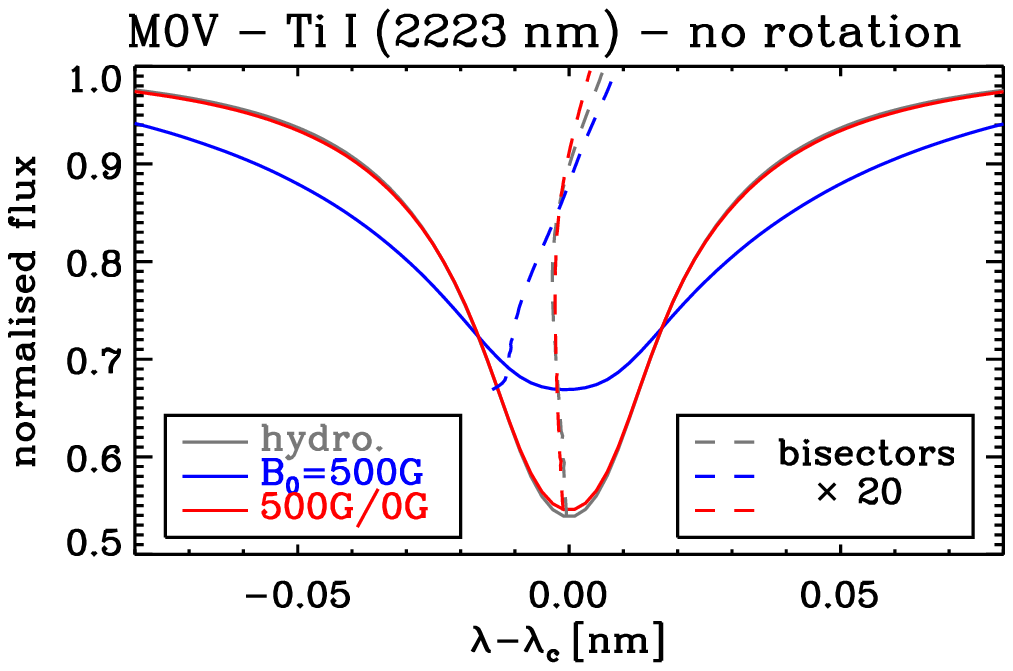}~\includegraphics[width=7.1cm]{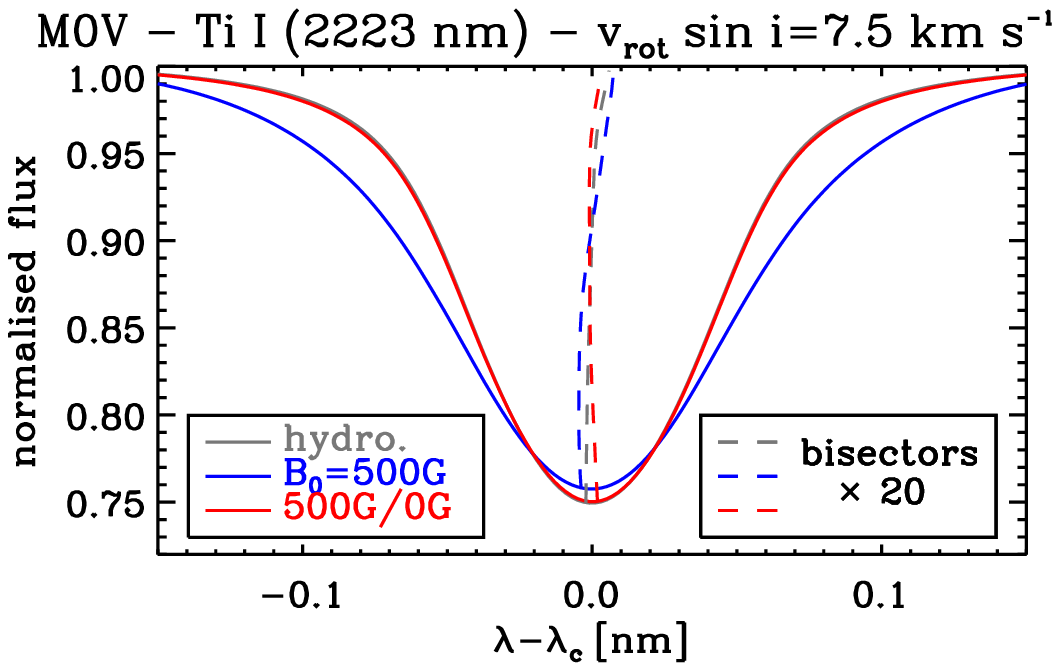}
\includegraphics[width=7.1cm]{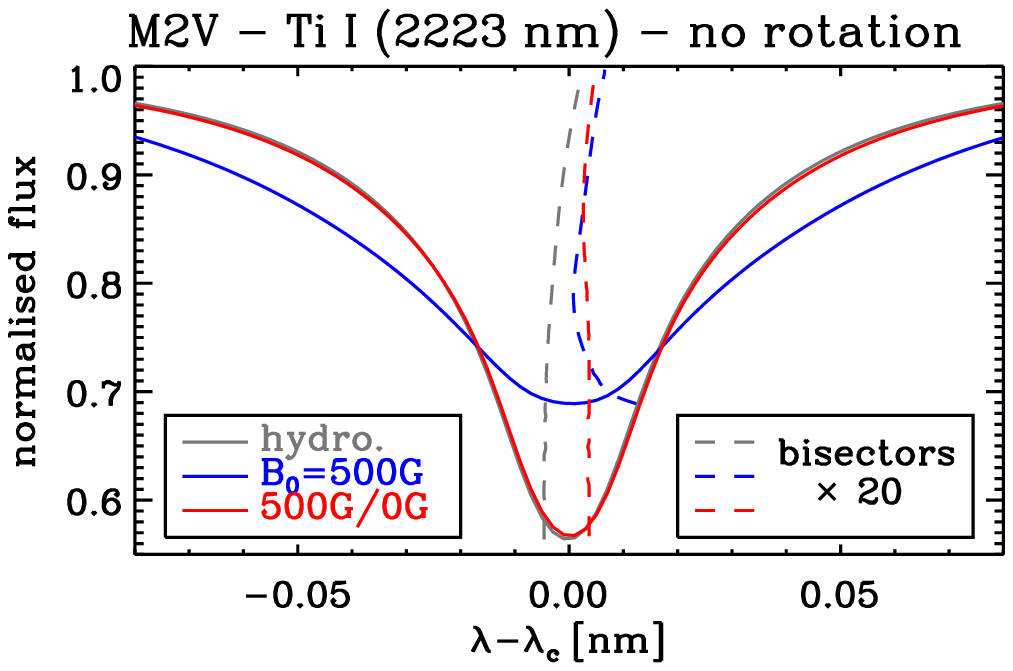}~\includegraphics[width=7.1cm]{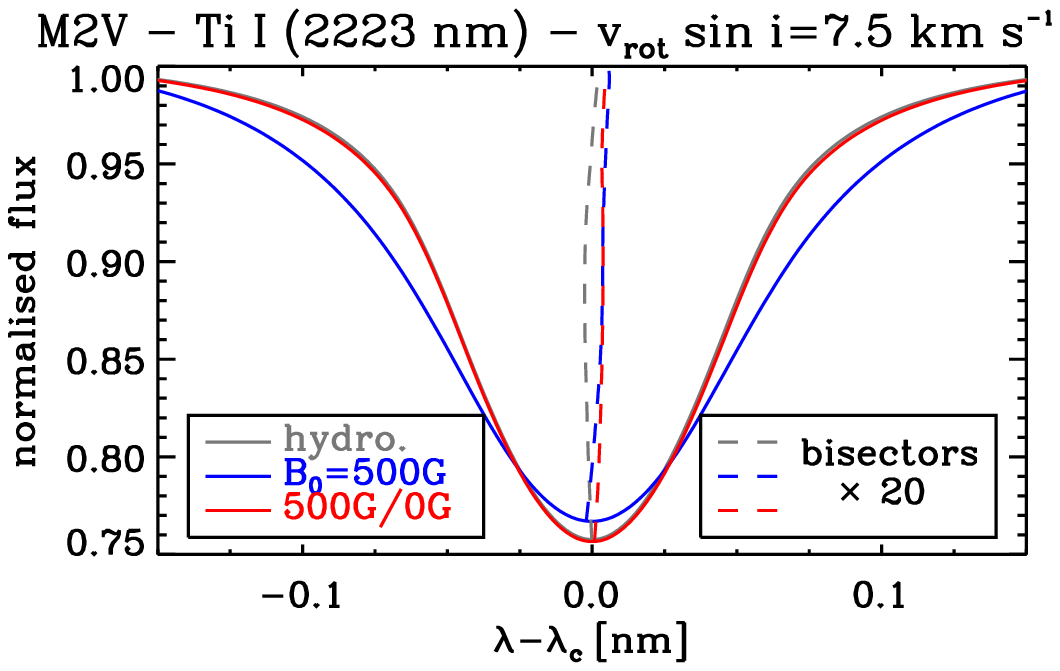}
\caption{Same as Fig.~\ref{fig:sample_app1}, but for the Ti\,\textsc{i} line at 2223\,nm and for the K- and M-star simulations.}\label{fig:sample_app2}
\end{figure*}
In Fig.~\ref{fig:LD_Johnson} in Sect.~\ref{sec:clv}, the magnetic limb brightening
is given in the Johnson passbands UX, B, V, R, and I for the G2V
star. Figure~\ref{fig:LD_app} shows the limb brightening for the F3V and M0V
simulations. The magnetic limb brightening is
largest for the F3V star and becomes smaller towards lower effective
temperatures as the faculae become less bright and less frequent (cf. Fig.~\ref{fig:inclview}).\par

In Fig.~\ref{fig:sample_TDvsZ} in Sect.~\ref{sec:discint}, disc-integrated
profiles of the Fe\,\textsc{i} line at 617.3\,nm are given for the G2V
simulations without magnetic field (grey curves) and with
$B_0=500\,\mathrm{G}$ (blue and red curves) where the magnetic field has been
ignored for the spectral line synthesis for the red curves in order to
disentangle the effect of the modified atmospheric structure from that of the
Zeeman effect. Here, we give analogous plots for further simulations for the
same spectral line (Fig.~\ref{fig:sample_app1}) and for the Ti\,\textsc{i}
line at 2223\,nm (Fig.~\ref{fig:sample_app2}).\par

As Fig.~\ref{fig:sample_app1} illustrates, the Fe\,\textsc{i} line is
significantly affected by line weakening in the F3V star. Similar to the
G2V-star case, the line is shifted to the red (in particular its wings) by the
modified convective flows. The Zeeman effect has a relatively small impact on
the spectral line profile. In the K stars, the thermodynamic modifications
seem less important than the Zeeman effect without rotation (left panels). In
the rotating case (right panel), both effects are of similar magnitude. As
they are of opposite sign, the apparent broadening of the line by the Zeeman
effect is reduced by the modifications in the atmosphere structure. In the M
stars (here, only the M0V is shown), the impact of the modified thermodynamics
and flows is very small. For these stars, systematic errors in Stokes-$I$
measurements of the magnetic field caused by differences in atmospheric
structure between magnetised and unmagnetised regions will probably be the
small.\par

Figure~\ref{fig:sample_app2} shows the same plots as
Fig.~\ref{fig:sample_app1} but for the Ti\,\textsc{i} line and only for the K-
and M-type stars. Although the line weakening is stronger for this line in the
K-type stars (cf. Fig.~\ref{fig:local_spectra}), possibly leading to an
underestimation of the unsigned average field from Stokes-$I$ measurements,
the general detection of magnetic field is probably more feasible in
this line as the broadening due to the Zeeman effect is larger and leads to
extended line wings. The line profiles from
the M-star simulations are almost unaffected by the modifications of the
thermodynamical atmosphere structure.\par
\end{document}